\documentclass[]{pasj00}

\begin{document}
\SetRunningHead{Sato et al.}{Substellar Companions to Seven Evolved
Intermediate-Mass Stars}
\Received{}
\Accepted{}

\title{Substellar Companions to Seven Evolved Intermediate-Mass Stars}



%
 \author{
   Bun'ei \textsc{Sato},\altaffilmark{1}
   Masashi \textsc{Omiya},\altaffilmark{1}
   Hiroki \textsc{Harakawa},\altaffilmark{1}
   Hideyuki \textsc{Izumiura},\altaffilmark{2,3}
   Eiji \textsc{Kambe},\altaffilmark{2}
   Yoichi \textsc{Takeda},\altaffilmark{3,4}
   Michitoshi \textsc{Yoshida},\altaffilmark{5}
   Yoichi \textsc{Itoh},\altaffilmark{6}
   Hiroyasu \textsc{Ando},\altaffilmark{3,4}
   Eiichiro \textsc{Kokubo}\altaffilmark{3,4}
   and
   Shigeru \textsc{Ida}\altaffilmark{1}
}
 \altaffiltext{1}{Department of Earth and Planetary Sciences, Tokyo Institute of Technology,
   2-12-1 Ookayama, Meguro-ku, Tokyo 152-8551, Japan}
 \email{satobn@geo.titech.ac.jp}
 \altaffiltext{2}{Okayama Astrophysical Observatory, National
   Astronomical Observatory of Japan, Kamogata,
   Okayama 719-0232, Japan}
 \altaffiltext{3}{The Graduate University for Advanced Studies,
   Shonan Village, Hayama, Kanagawa 240-0193, Japan}
 \altaffiltext{4}{National Astronomical Observatory of Japan, 2-21-1 Osawa,
   Mitaka, Tokyo 181-8588, Japan}
 \altaffiltext{5}{Hiroshima Astrophysical Science Center, Hiroshima University,
   Higashi-Hiroshima, Hiroshima 739-8526, Japan}
 \altaffiltext{6}{Graduate School of Science, Kobe University,
   1-1 Rokkodai, Nada, Kobe 657-8501, Japan}
 

\maketitle

\begin{abstract}
We report the detections of substellar companions orbiting around seven evolved
intermediate-mass stars from precise Doppler measurements at Okayama
Astrophysical Observatory.
$o$ UMa (G4 II-III) is a giant with a mass of 3.1 $M_{\odot}$ and hosts
a planet with minimum mass of $m_2\sin i=4.1~M_{\rm J}$ in an orbit
with a period $P=1630$ d and an eccentricity $e=0.13$.
This is the first planet candidate ($< 13~M_{\rm J}$) ever discovered
around stars more massive than $3~M_{\odot}$.
$o$ CrB (K0 III) is a 2.1 $M_{\odot}$ giant and has a planet
of $m_2\sin i=1.5~M_{\rm J}$ in a 187.8 d orbit with $e=0.19$.
This is one of the least massive planets ever discovered
around $\sim2~M_{\odot}$ stars.
HD 5608 (K0 IV) is an 1.6 $M_{\odot}$ subgiant
hosting a planet of $m_2\sin i=1.4~M_{\rm J}$ in a 793 d orbit with $e=0.19$.
The star also exhibits a linear velocity trend suggesting the existence of an
outer, more massive companion.
75 Cet (G3 III:) is a 2.5 $M_{\odot}$ giant hosting a planet
of $m_2\sin i=3.0~M_{\rm J}$ in a 692 d orbit with $e=0.12$.
The star also shows possible additional periodicity of about 200 d
and 1880 d with velocity amplitude of $\sim7$--10 m s$^{-1}$, although
these are not significant at this stage.
$\nu$ Oph (K0 III) is a 3.0 $M_{\odot}$ giant and has two brown-dwarf
companions of $m_2\sin i= 24~M_{\rm J}$ and 27 $M_{\rm J}$,
in orbits with $P=530.3$ d and 3190 d, and $e=0.126$ and 0.17, respectively,
which were independently announced by \citet{quirrenbach:2011}.
The ratio of the periods is close to 1:6, suggesting that the companions
are in mean motion resonance.
We also independently confirmed planets around $\kappa$ CrB (K0 III-IV) and
HD 210702 (K1 IV), which had been announced by \citet{johnson:2008}
and \citet{johnson:2007a}, respectively. All of the orbital parameters
we obtained are consistent with the previous results.
\end{abstract}

\section{Introduction}\label{intro}
Intermediate-mass (1.5--5 $M_{\odot}$) stars have been gathering
more attention of researchers as promising sites of planet formation.
It is not only because some planetary systems were found around A-type
dwarfs by direct imaging (e.g. \cite{marois:2008}) but also intensive Doppler
surveys of intermediate-mass giants and subgiants, which are
``evolved counterparts of A-type dwarfs'', have unveiled remarkable
properties of planets around them
(e.g. \cite{frink:2002}; \cite{setiawan:2005}; \cite{sato:2008b}; \cite{sato:2010};
\cite{hatzes:2005}; \cite{hatzes:2006}; \cite{johnson:2011a}; \cite{lovis:2007};
\cite{niedzielski:2009b}; \cite{dollinger:2009}; \cite{demedeiros:2009};
\cite{wang:2011}; \cite{omiya:2011}; \cite{wittenmyer:2011}).
About 50 substellar companions have been found around such evolved
intermediate-mass stars, whose masses, semimajor axes, and eccentricities
are ranging from 0.6 to 40 $M_{\rm J}$, from 0.08 to 6 AU, and from 0 to 0.5,
respectively\footnote{see e.g. http://exoplanet.eu}.
The occurrence rate of giant planets increases as stellar mass at least
up to $\sim$1.9$M_{\odot}$ ($\sim$10--20\%; e.g. \cite{johnson:2007b};
\cite{bowler:2010}), and super-massive ($\gtrsim5M_{\rm J}$) planets are
more abundant around more massive ($\gtrsim 2M_{\odot}$) giants
(e.g. \cite{lovis:2007}).
Positive correlation between frequency of giant planets and stellar
metallicity may exist in subgiants like in solar-type stars
(\cite{johnson:2010}) but not be seen in giants
(e.g. \cite{pasquini:2007}; \cite{takeda:2008}).
Semimajor-axis distribution of the planets is one of the most
interesting features; almost all the planets found around intermediate-mass
evolved stars reside in orbits with semimajor axis larger than 0.6 AU
(e.g. \cite{johnson:2007a}; \cite{sato:2008a}), while many short-period planets
exist around low-mass FGK dwarfs.
Multi-planet systems have also been found around evolved intermediate-mass
stars including candidates in mean motion resonance
(MMR; \cite{niedzielski:2009a}; \cite{johnson:2011b}; \cite{quirrenbach:2011}).
All the properties give us deep insight into formation and evolution of
substellar companions around intermediate-mass stars.

We here report the detections of new substellar companions around
evolved intermediate-mass stars emerged from the Okayama Planet
Search Program (\cite{sato:2005}). The program has been regularly
monitoring radial velocities (RVs) of about 300 intermediate-mass GK
giants since 2001 at Okayama Astrophysical Observatory (OAO) and
announced 9 planets and 1 brown dwarf so far
(\cite{sato:2003}; \cite{sato:2007}; \cite{sato:2008a}; \cite{sato:2008b};
\cite{liu:2008}).
We also discovered 3 planets and 3 brown-dwarf candidates
around GK giants using 2.16m telescope at Xinglong observatory in
China, 1.8m telescope at Bohyunsan Optical Astronomy Observatory
in Korea, and Subaru 8.2m telescope in Hawaii as well as OAO 1.88m
telescope under the framework of East-Asian Planet Search Network
(\cite{liu:2009}; \cite{omiya:2009}; \cite{omiya:2011};
\cite{sato:2010}; \cite{wang:2011}).

Rest of the paper is organized as follows. We describe the observations
in section \ref{obs} and the stellar properties are presented in section \ref{stpara}.
Analyses of RV, linear trend, period search, orbital solution,
and line shape variation are described in section \ref{ana} and the results are
presented in section \ref{results}.
Section \ref{summary} is devoted to summary and discussion.

\section{Observation}\label{obs}
All of the observations were made with the 1.88 m telescope and the HIgh
Dispersion Echelle Spectrograph (HIDES; \cite{izumiura:1999}) at OAO.
In 2007 December, HIDES was upgraded from single CCD (2K$\times$4K)
to a mosaic of three CCDs. The upgrade enabled us to obtain spectra
covering a wavelength range of 3750--7500${\rm \AA}$ using a red
cross-disperser, which is about three times wider than before the upgrade.

For precise RV measurements, we used an iodine absorption
cell (I$_2$ cell; \cite{kambe:2002}), which provides a fiducial
wavelength reference in a wavelength range of 5000--5800${\rm \AA}$
(covered by the middle CCD after the upgrade in 2007 December).
A slit width is set to 200 $\mu$m ($0.76^{\prime\prime}$)
giving a spectral resolution ($R=\lambda/\Delta\lambda$) of 67000
by about 3.3 pixels sampling. We can typically obtain a
signal-to-noise ratio S/N$>$200 pix$^{-1}$ for a $V<6.5$
star with an exposure time shorter than 30 min.
The reduction of echelle data (i.e. bias subtraction, flat-fielding,
scattered-light subtraction, and spectrum extraction) is performed
using the IRAF\footnote{IRAF is distributed by the National
Optical Astronomy Observatories, which is operated by the
Association of Universities for Research in Astronomy, Inc. under
cooperative agreement with the National Science Foundation,
USA.} software package in the standard way.

\section{Stellar Properties}\label{stpara}
\citet{takeda:2008} derived atmospheric parameters
(effective temperature $T_{\rm eff}$, surface gravity $\log g$,
micro-turbulent velocity $v_t$, and Fe abundance [Fe/H])
of all the targets for Okayama Planet Search Program based on the spectroscopic
approach using the equivalent widths of well-behaved Fe I and Fe II
lines of iodine-free stellar spectra. Details of the procedure and resultant
parameters are presented in \citet{takeda:2002} and \citet{takeda:2008}.

\begin{figure}
  \begin{center}
    \FigureFile(85mm,80mm){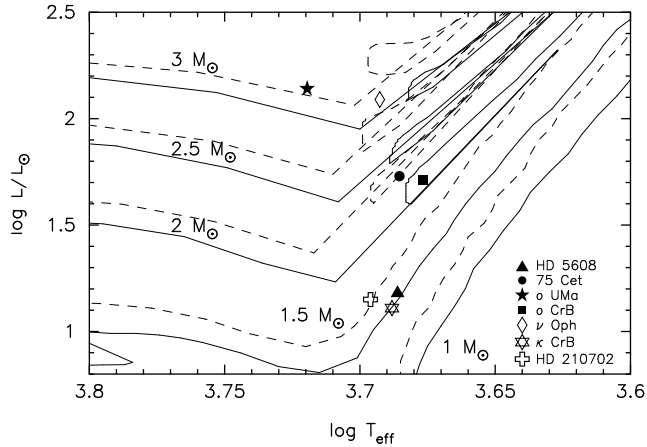}
  \end{center}
\caption{HR diagram of the planet-harboring stars presented in this paper.
Pairs of evolutionary tracks from Lejeune and Schaerer (2001)
for stars with $Z=0.02$ (solar metallicity; solid
lines) and $Z=0.008$ (dashed lines) of masses between 1 and 3
$M_{\odot}$ are also shown.}\label{fig-HRD}
\end{figure}

They also determined the absolute magnitude $M_V$ of each star
from the apparent $V$-band magnitude and Hipparcos
parallax $\pi$ (\cite{esa:1997}) taking account of
correction of interstellar extinction $A_V$ based on \citet{arenou:1992}'s
table. The bolometric correction $B.C.$ was estimated based on the
\citet{kurucz:1993}'s theoretical calculation. With use of these parameters
and theoretical evolutionary tracks of \citet{lejeune:2001},
the luminosity $L$ and mass $M$ of each star were obtained.
The stellar radius $R$ was estimated using the Stefan-Boltzmann
relationship and the measured $L$ and $T_{\rm eff}$.  The stars presented
herein are plotted on the HR diagram in Figure \ref{fig-HRD} and
their properties are summarized in Table \ref{tbl-stars}.

The uncertainties for atmospheric and physical parameters of the stars
were also derived by \citet{takeda:2008} and we quoted the values
in Table \ref{tbl-stars}. Since details of the procedure are described in
\citet{takeda:2008}, we here briefly summarize it.
The uncertainties of [Fe/H], $T_{\rm eff}$, $\log g$, and $v_{\rm t}$ presented
in the table are
internal statistical errors for a given data set of Fe~{\sc i} and Fe~{\sc ii}
line equivalent widths (see subsection 5.2 of \cite{takeda:2002}).
Since these parameter values are sensitive to slight changes in the
equivalent widths as well as to the adopted set of lines (\cite{takeda:2008}),
realistic uncertainties may be by a factor of $\sim$ 2--3 larger
than these estimates from a conservative point of view
(e.g., 50--100 $K$ in $T_{\rm eff}$, 0.1--0.2~dex in $\log g$).
Therefore the ranges of stellar mass were obtained by perturbing the
values of $\log L$, $\log T_{\rm eff}$, and $\rm [Fe/H]$ interchangeably
by typical amounts of uncertainties; $\Delta\log L$ corresponding to parallax
errors given in the Hipparcos catalog, $\Delta\log T_{\rm eff}$ of $\pm0.01$ dex
almost corresponding to $\sim\pm100$~K, and $\Delta{\rm [Fe/H]}$ of $\pm0.1$ dex.
Similarly, the error in $R$ was evaluated from $\Delta\log L$ and
$\Delta\log T_{\rm eff}$ (see the section 3.2  and footnote 8 of \cite{takeda:2008}
for the details of the procedure).
The resulting mass value may also appreciably depend on the chosen
set of theoretical evolutionary tracks (e.g., the systematic
difference as large as $\sim 0.5M_{\odot}$ for the case of
metal-poor tracks between \cite{lejeune:2001} and
\cite{girardi:2000}.; see also footnote 3 in \cite{sato:2008a}).

\begin{figure}
  \begin{center}
    \FigureFile(85mm,80mm){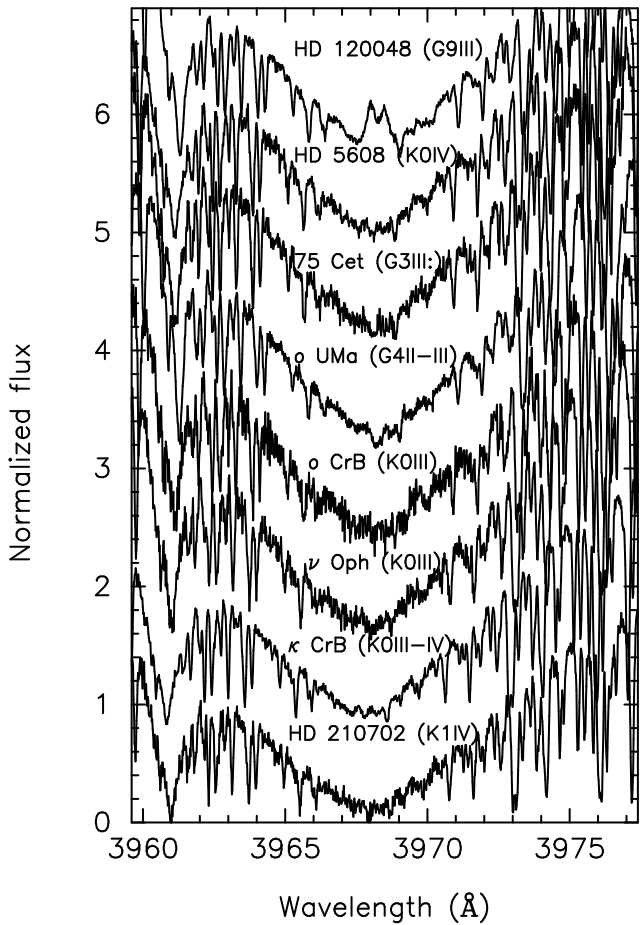}
  \end{center}
\caption{Spectra in the region of Ca H lines. All of the stars show no
significant core reversals in the lines compared to that in the
chromospheric active star HD 120048, which shows velocity scatter
of about 20 m s$^{-1}$.
A vertical offset of about 0.8 is added to each spectrum.}\label{fig-CaH}
\end{figure}

Hipparcos observations revealed photometric stability for the stars
down to $\sigma_{\rm HIP}=0.004-0.008$ mag. Furthermore, all the stars
show no significant emission in the core of Ca II HK lines as shown in
Figure \ref{fig-CaH}, suggesting that the stars are chromospherically inactive.

\section{Analysis}\label{ana}
\subsection{Radial Velocity Analysis}\label{rv ana}
For RV analysis, we basically adopted the modeling technique of an I$_2$-superposed
stellar spectrum (star+I$_2$) detailed in \citet{sato:2002}, which
is based on the method by \citet{butler:1996}. In the technique,
a star+I$_2$ spectrum is modeled as a product of a high resolution I$_2$ and
a stellar template spectrum convolved with a modeled instrumental profile
(IP) of the spectrograph. To obtain the stellar template, \citet{sato:2002}
extracted a high resolution stellar spectrum from several star+I$_2$
spectra. However, we here used a stellar template that was obtained by
deconvolving a pure stellar spectrum with the spectrograph IP estimated
from an I$_2$-superposed B-star or Flat spectrum because we finally
found that the overall precision in RV using the template thus
obtained was slightly better than that based on the technique in
\citet{sato:2002}. We have now achieved a long-term Doppler precision
of about 4--5 m s$^{-1}$ over a time span of 9 years.
Measurement error was estimated from an ensemble of velocities from each
of $\sim$300 spectral regions (each $\sim$3${\rm \AA}$ long) in every exposure. 

\subsection{Linear Trend}\label{linear trend}
We examined whether there is evidence of a linear trend in RVs
with timescale much larger than the duration of observations.  To address
the significance of the velocity trend, we used the $F$-test in a similar way
to that described by \citet{cumming:1999}. The weighted sum of squares
of residuals to the best-fit linear slope $\chi_{N-2}^2$ is compared with the
weighted sum of squares about the mean $\chi_{N-1}^2$. If there is no
long-term trend in observed velocities and the residual follow Gaussian
distribution, the statistic
\begin{eqnarray}\label{F-test}
F &=& (N-2)\frac{\chi_{N-1}^2-\chi_{N-2}^2}{\chi_{N-2}^2}
\end{eqnarray}
follows Fisher's $F$ distribution with 1 and $N-2$ degrees of freedom which
measures how much the fit is improved by introducing the linear trend. Using
this probability distribution, we can estimate the False Alarm Probability $(FAP)$
that pure noise fluctuations would produce a linear velocity trend by chance.
This method, however, assumes that the errors follow Gaussian distribution,
which is not necessarily the case for the actual observations.
In stead we here adopt so called a bootstrap randomization method to estimate
$FAP$. In this approach, the observed RVs are randomly redistributed, keeping fixed
the observation time. The major advantage of this method is that one can derive
$FAP$ without assuming any particular distribution of noise. For each star,
we generated 10$^5$ fake datasets in this way, applied the same analysis
to them, and obtained $F$-value by the equation (\ref{F-test}). The frequency of
fake datasets whose $F$ exceeded the observed one was adopted as a $FAP$
for the trend.

\subsection{Period Search}\label{period search}
To search for periodicity in RV data we performed a Lomb-Scargle periodogram
analysis (\cite{scargle:1982}). To assess the significance of this periodicity, we estimate
False Alarm Probability ($FAP$), using a bootstrap randomization method in which
the observed RVs were randomly redistributed, keeping fixed
the observation time. We generated 10$^5$ fake datasets in this way,
and applied the same periodogram analysis to them. The frequency of fake datasets
which exhibit a periodogram power higher than the observed dataset is
adopted as a $FAP$ for the signal.

\subsection{Orbital Solution}\label{orbit}
The best-fit Keplerian orbit for the data was derived using a Levenberg-Marquardt
fitting algorithm (\cite{press:1989}) to obtain a minimum chi-squared
solution by varying the free parameters (orbital period $P$,
time of periastron passage $T_p$, eccentricity $e$, velocity amplitude $K_1$
and argument of periastron $\omega$ for each companion).
When we found a significant linear trend in observed
RVs by the analysis described in section \ref{linear trend}, we included the
slope as a free parameter and simultaneously derived the best-fit trend
and Keplerian orbit for the data.

Each star has stellar ``jitter'', which is intrinsic variability
in RV as a source of astrophysical noise such as stellar
oscillation and chromospheric activity, and also unknown systematic
measurement error. To account for the variability,
we have quadratically added a jitter $\sigma_{\rm jitter}$
to the RV measurement uncertainties before performing the final
least-squared fitting of the orbit so that
the resultant reduced $\chi^2$ would become unity.

The uncertainty for each orbital parameter was estimated using a bootstrap
Monte Carlo approach, subtracting the theoretical fit, scrambling the residuals,
adding the theoretical fit back to the residuals and then refitting.

\subsection{Line Shape Analysis}\label{line shape}
We performed spectral line shape analysis based on high resolution
stellar templates to investigate other possible causes of
apparent RV variations such as pulsation
and rotational modulation rather than orbital motion.
Details of the analysis are described in \citet{sato:2007} and
\citet{sato:2002}, and here we briefly summarize the procedure.

At first, we extracted two stellar templates from five star+I$_2$ spectra
at nearly the peak and valley phases of observed RVs for
each star. Cross correlation profiles of the two templates
were then calculated for about 40--100 spectral segments (4--5${\rm \AA}$
width each) in which severely blended lines or broad lines were
not included.
We calculated three bisector quantities for the derived
cross correlation profile for each segment:
the velocity span (BVS), which is the velocity difference
between two flux levels of the bisector;
the velocity curvature (BVC), which is the difference of the
velocity span of the upper half and lower half of the bisector;
and the velocity displacement (BVD), which is the average of
the bisector at three different flux levels.
Flux levels of 25\%, 50\%, and 75\% of the cross
correlation profile were used to calculate the above three
bisector quantities.
If both of the BVS and the BVC for stars are identical to zero and the
average BVD agrees with the velocity difference between the two
templates at the peak and valley phases of observed RVs
($\simeq 2K_1$), the cross correlation profiles can be considered to be
symmetric and thus the observed RV variations are considered
to be due to parallel shifts of the spectral lines, which is consistent with
the planetary hypothesis.

\section{Results}\label{results}

\subsection{HD 5608 (HIP 4552)}\label{HD5608}

\begin{figure}
  \begin{center}
    \FigureFile(85mm,80mm){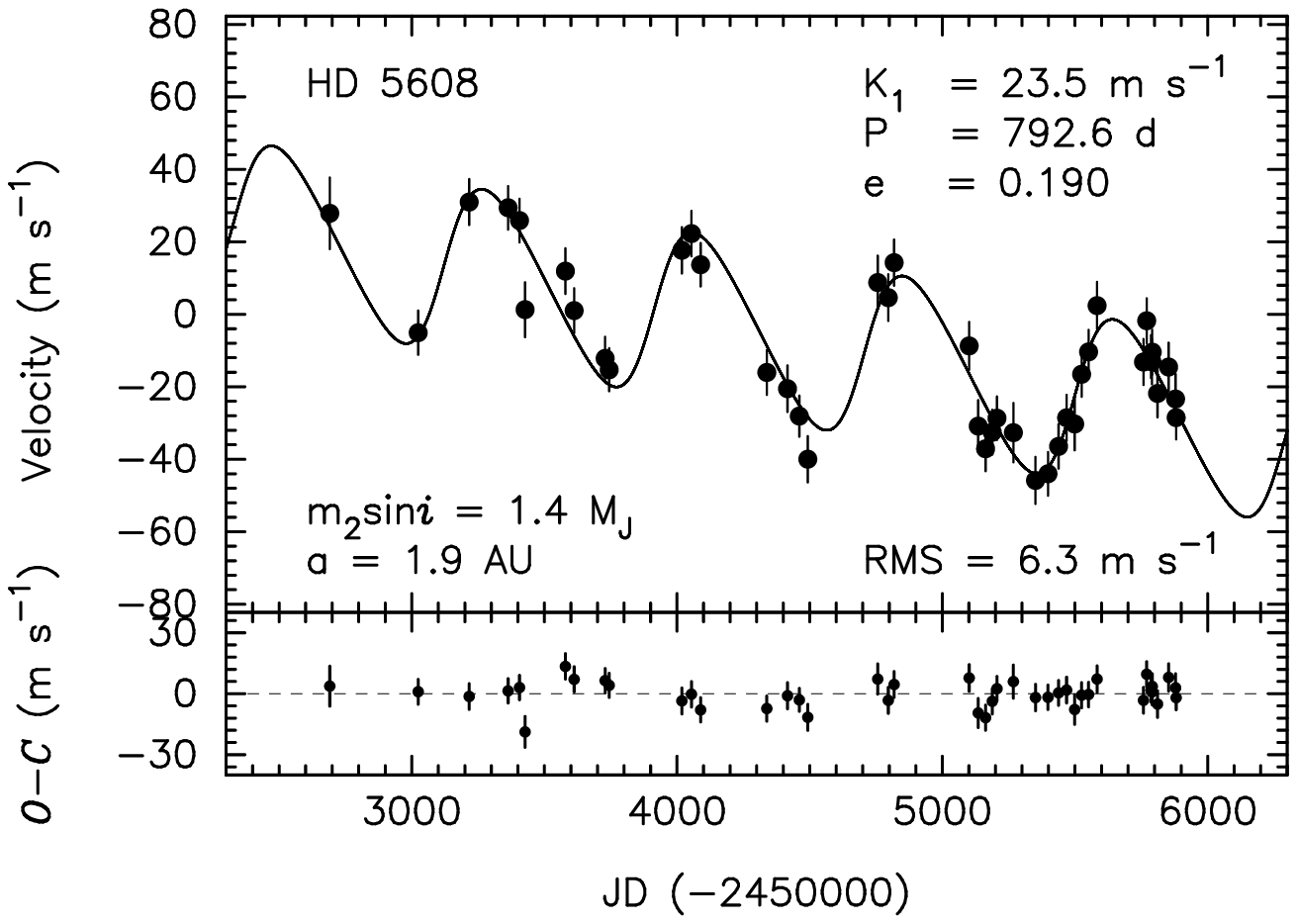}
  \end{center}
\caption{{\it Top}: Radial velocities of HD 5608 observed at OAO.
The Keplerian orbit with the linear trend is shown by the solid line.
The error bar for each point includes the stellar jitter estimated in
section \ref{HD5608}.
{\it Bottom}: Residuals to the orbital fit.
The rms to the fit is 6.3 m s$^{-1}$.}
\label{fig-HD5608}
\end{figure}

We collected a total of 43 RV data of HD 5608 between
2003 February and 2011 November. 
The observed RVs are shown in Figure \ref{fig-HD5608}
and are listed in Table \ref{tbl-HD5608} together with their
estimated uncertainties.

Based on the linear-trend test described in section \ref{linear trend},
the star showed a significant linear trend of $-4.9$ m s$^{-1}$ yr$^{-1}$
with $FAP=5\times10^{-5}$, suggesting the existence of an distant companion.
After removing the trend from the observed RVs, Lomb-Scargle periodogram
showed a significant peak around 766 day with a $FAP<10^{-5}$.
Simultaneous fitting of a single Keplerian model and a linear trend to
the data yielded an orbital period of $P=792.6\pm7.7$ days,
a velocity semiamplitude $K_1=23.5\pm1.6$ m s$^{-1}$, and an eccentricity
$e=0.190\pm0.061$ for the inner companion, and the acceleration of
$\dot{\gamma}=-5.5$ m s$^{-1}$ yr$^{-1}$
for the linear trend. When we adopt the stellar mass of $1.55~M_{\odot}$ for the
star, we obtained a minimum mass of the inner companion
$m_2\sin i=1.4~M_{\rm J}$ and a semimajor axis $a=1.9$ AU.
For the outer companion, we may give an order-of-magnitude relation
between $\dot{\gamma}$ and the properties of the companion (e.g. \cite{winn:2009})
\begin{eqnarray}\label{HD5608c}
\frac{m_c \sin i_c}{a_c^2} &\sim& \frac{\dot{\gamma}}{G} =
(0.031\pm0.003)~M_{\rm J}~{\rm AU^{-2}}
\end{eqnarray}
where $m_c$ is the companion mass, $i_c$ is the orbital inclination, and
$a_c$ is the orbital radius.

The rms scatter of the residuals to the Keplerian fit was
6.3 m s$^{-1}$.
The adopted stellar jitter for the star (see section \ref{orbit}) is
$\sigma_{\rm jitter}=5.0$ m s$^{-1}$, which is consistent with typical
one for subgiants (\cite{johnson:2008}), and we found no significant
periodicity in the residuals.
The resulting model is shown in Figure \ref{fig-HD5608}
overplotted on the velocities, whose error bars include the stellar jitter,
and its parameters are listed in Table \ref{tbl-planets}.

For line shape analysis, we extracted templates with velocity of
$\sim$10 m s$^{-1}$ and $\sim$$-$20--$-$30 m s$^{-1}$, and calculated
cross correlation profile between them. As seen in Table \ref{tbl-bisector},
resultant BVS and BVC values of the cross correlation profiles
did not show any significant variability, and BVD value agreed
with the velocity difference between the two templates,
which are consistent with the planetary hypothesis as the
cause of the observed RV variations.

\subsection{75 Cet (HD 15779, HIP 11791)}\label{75Cet}

\begin{figure}
  \begin{center}
    \FigureFile(85mm,80mm){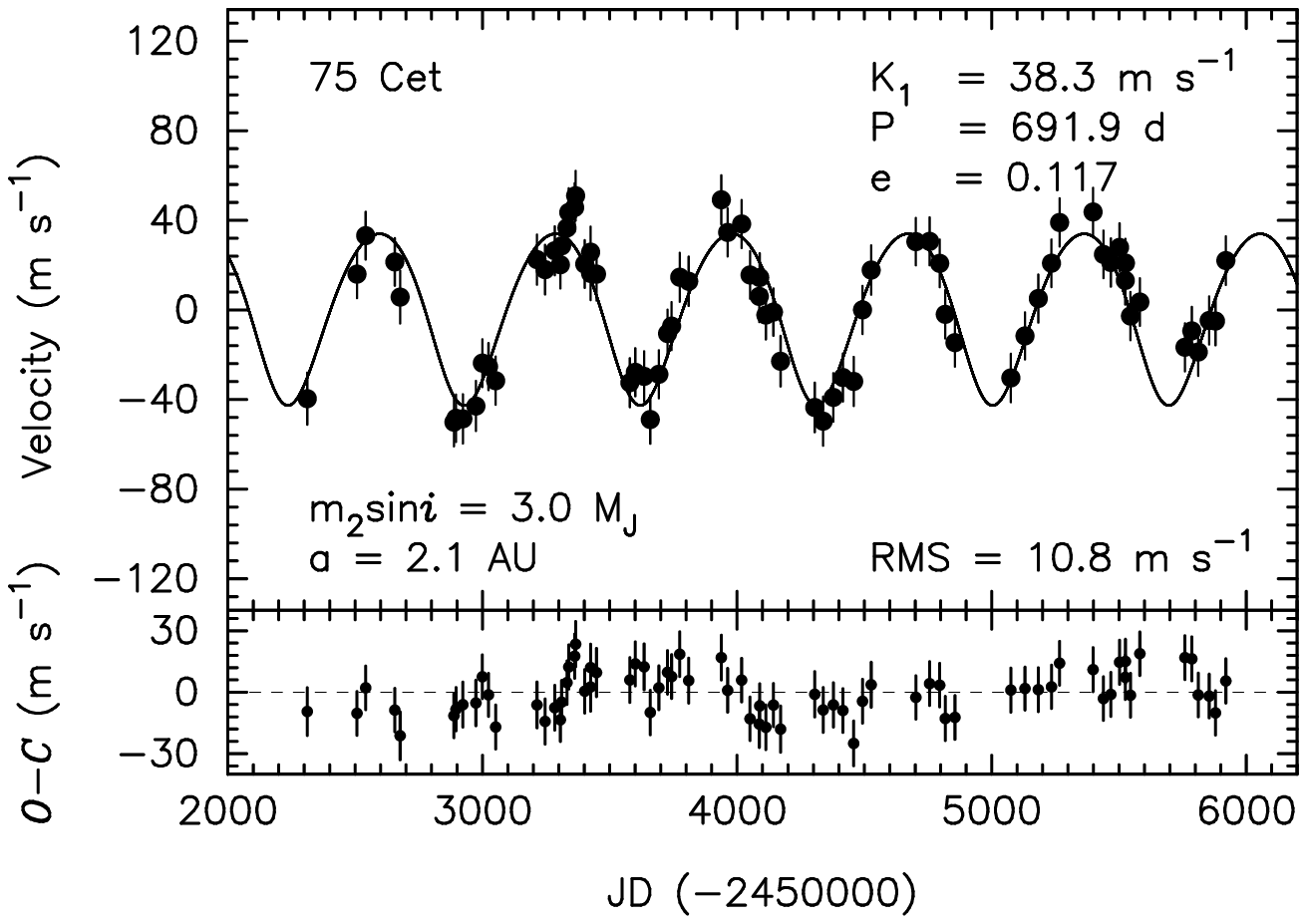}
  \end{center}
\caption{{\it Top}: Radial velocities of 75 Cet observed at OAO.
The Keplerian orbit is shown by the solid line.
The error bar for each point includes the stellar jitter estimated in
section \ref{75Cet}.
{\it Bottom}: Residuals to the Keplerian fit.
The rms to the fit is 10.8 m s$^{-1}$.}
\label{fig-HD15779}
\end{figure}

We collected a total of 74 RV data of 75 Cet between
2002 February and 2011 December.
The observed RVs are shown in Figure \ref{fig-HD15779}
and are listed in Table \ref{tbl-HD15779} together with their
estimated uncertainties. Lomb-Scargle periodogram of the data exhibits
a dominant peak at a period of 698 days with a $FAP<1\times10^{-5}$.

Single Keplerian model for the star yielded orbital parameters for
the companion of $P=691.9\pm3.6$ days, $K_1=38.3\pm2.0$ m s$^{-1}$,
and $e=0.117\pm0.048$. Adopting a stellar mass of
2.49 $M_{\odot}$, we obtain $m_2\sin i=3.0~M_{\rm J}$ and $a=2.1$ AU
for the companion. The rms scatter of the residuals to the Keplerian
fit is 10.8 m s$^{-1}$ and the reduced $\sqrt{\chi^2}$ was 2.7 when we
adopted the measurement errors as weight of the least-squared fitting.

\begin{figure}
  \begin{center}
    \FigureFile(85mm,80mm){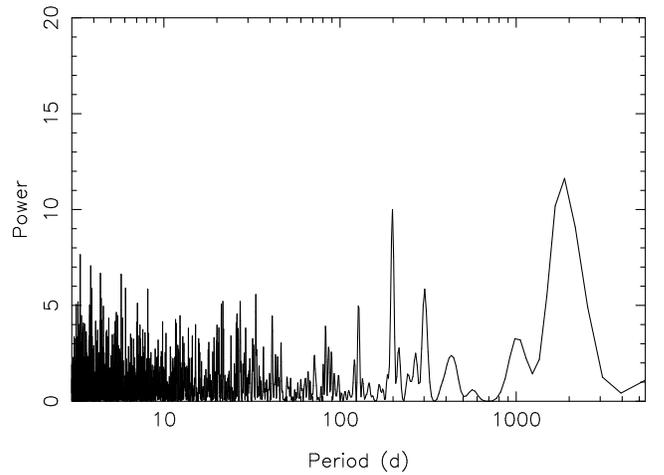}
  \end{center}
\caption{Periodogram of the residuals to the Keplerian fit
for 75 Cet. Possible peaks ($FAP\sim0.003$) are seen at
periods of about 200 and 1880 days.}
\label{fig-periodogram}
\end{figure}

We performed periodogram analysis to the residuals and found
possible peaks around 200 d and 1880 d (Figure \ref{fig-periodogram}).
Although these periodicity is not significant at this stage
($FAP\sim3\times10^{-3}$), Keplerian fitting yielded velocity
semiamplitudes of $\sim7$ m s$^{-1}$ and $\sim10$ m s$^{-1}$ for the
periodicity, respectively, which corresponds to
$m_2\sin i\sim0.4~M_{\rm J}$ and $1~M_{\rm J}$,
and $a\sim0.9$ AU and $\sim$4 AU, respectively.
Continuous monitoring of the star will enable us to validate
the periodicity.
We here have quadratically added a jitter of 10 m s$^{-1}$ to the
RV uncertainties to account for the RMS scatter to the single
Keplerian model. The resulting model is shown in Figure \ref{fig-HD15779}
overplotted on the velocities, and its parameters are listed in
Table \ref{tbl-planets}.

We did not find any significant line-shape variability corresponding
to the primary period of 692 d (Table \ref{tbl-bisector}),
while those corresponding to the other possible periods
were below the detection limit of the line shape analysis
because of their low velocity amplitudes. 

\subsection{o UMa (HD 71369, HIP 41704)}\label{oUMa}

\begin{figure}
  \begin{center}
    \FigureFile(85mm,80mm){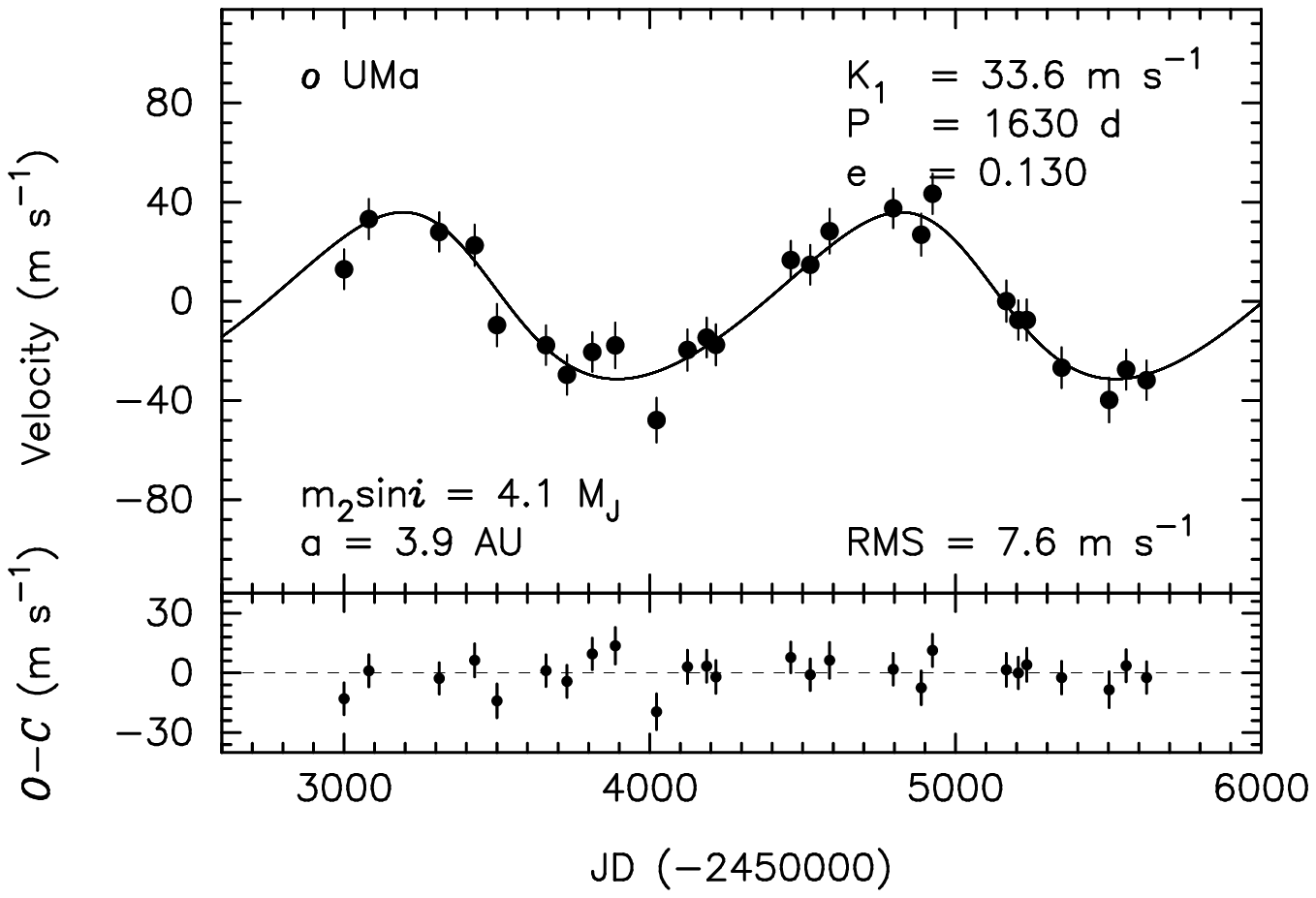}
  \end{center}
\caption{{\it Top}: Radial velocities of $o$ UMa observed at OAO.
The Keplerian orbit is shown by the solid line.
The error bar for each point includes the stellar jitter estimated in
section \ref{oUMa}.
{\it Bottom}: Residuals to the Keplerian fit.
The rms to the fit is 7.6 m s$^{-1}$.}
\label{fig-HD71369}
\end{figure}

We collected a total of 26 RV data of $o$ UMa between
2003 December and 2011 March.
The observed RVs are shown in Figure \ref{fig-HD71369}
and are listed in Table \ref{tbl-HD71369} together with their
estimated uncertainties. Lomb-Scargle periodogram of the data exhibits
a dominant peak at a period of 1575 days with a $FAP<1\times10^{-5}$.

Single Keplerian model for the star yielded orbital parameters for
the companion of $P=1630\pm35$ days, $K_1=33.6\pm2.1$ m s$^{-1}$,
and $e=0.130\pm0.065$. Adopting a stellar mass of
3.09 $M_{\odot}$, we obtain $m_2\sin i=4.1~M_{\rm J}$ and $a=3.9$ AU
for the companion.
The rms scatter of the residuals to the Keplerian fit was
7.6 m s$^{-1}$. We found no significant periodicity in the
residuals and then we adopted a stellar jitter
$\sigma_{\rm jitter}=6.5$ m s$^{-1}$ for the star.
The resulting model is shown in Figure \ref{fig-HD71369}
overplotted on the velocities, and its parameters are listed in
Table \ref{tbl-planets}. We did not find any significant line
shape variability for the star corresponding to the observed
RV variations (Table \ref{tbl-bisector}).

\subsection{o CrB (HD 136512, HIP 75049)}\label{oCrB}

\begin{figure}
  \begin{center}
    \FigureFile(85mm,80mm){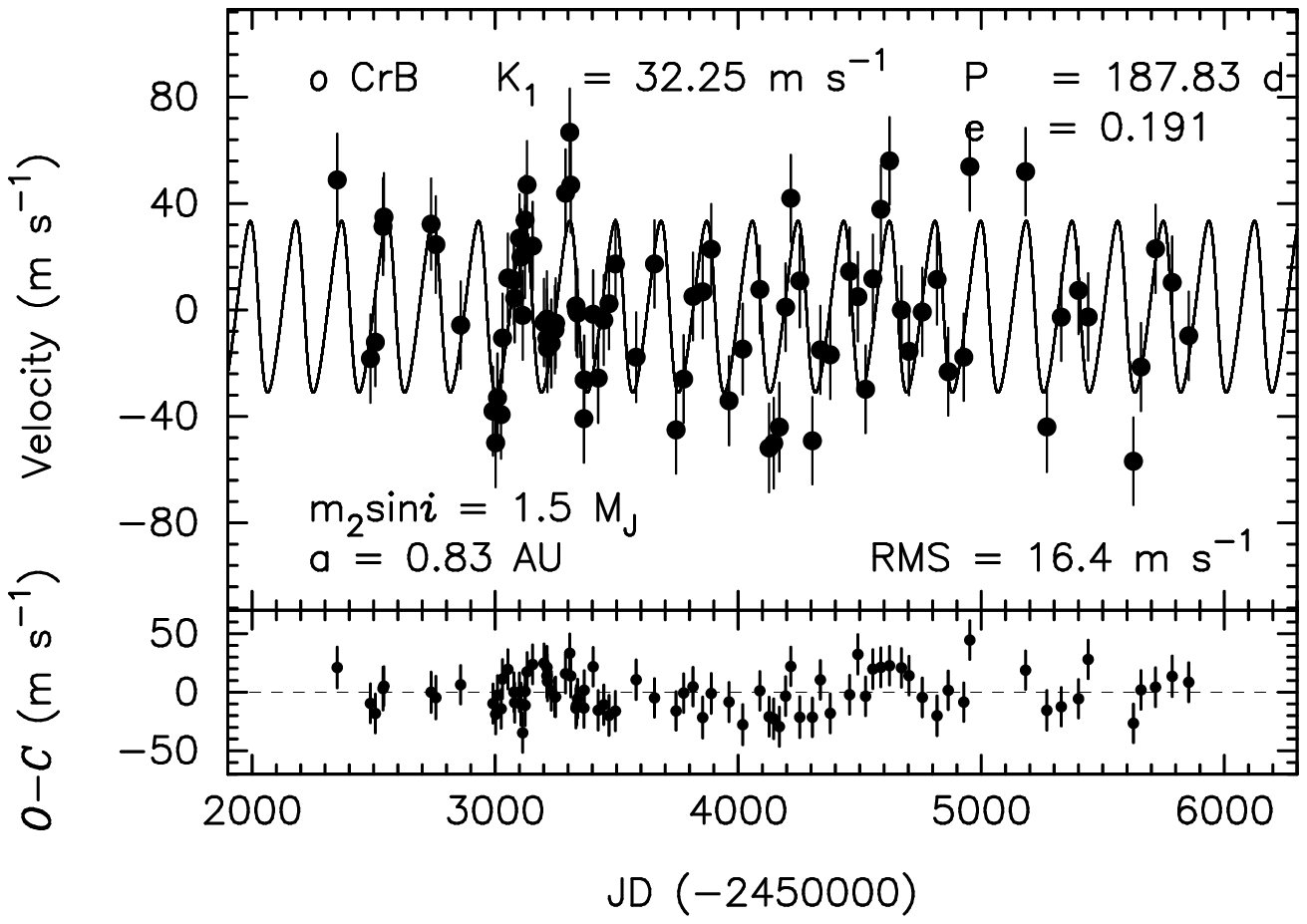}
  \end{center}
\caption{{\it Top}: Radial velocities of $o$ CrB observed at OAO.
The Keplerian orbit is shown by the solid line.
The error bar for each point includes the stellar jitter estimated in
section \ref{oCrB}.
{\it Bottom}: Residuals to the Keplerian fit.
The rms to the fit is 16.4 m s$^{-1}$.}
\label{fig-HD136512}
\end{figure}

We collected a total of 85 RV data of $o$ CrB between
2002 March and 2011 October.
The observed radial velocities are shown in Figure \ref{fig-HD136512}
and are listed in Table \ref{tbl-HD136512} together with their
estimated uncertainties. Lomb-Scargle periodogram of the data exhibits
a dominant peak at a period of 187.7 days with a $FAP<1\times10^{-5}$.

\begin{figure}
  \begin{center}
    \FigureFile(85mm,80mm){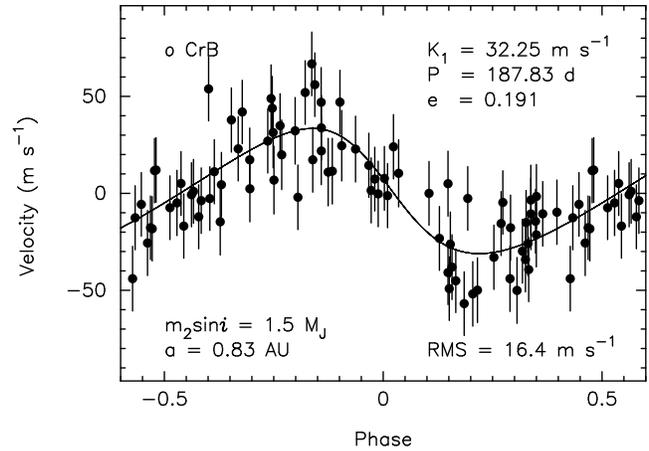}
  \end{center}
\caption{Phased radial velocities of $o$ CrB observed at OAO.
The Keplerian orbit is shown by the solid line.
The error bar for each point includes the stellar jitter estimated in
section \ref{oCrB}.}
\label{fig-HD136512p}
\end{figure}

Single Keplerian model for the star yielded orbital parameters for
the companion of $P=187.83\pm0.54$ days, $K_1=32.25\pm2.8$ m s$^{-1}$,
and $e=0.191\pm0.085$. Adopting a stellar mass of
2.13 $M_{\odot}$, we obtain $m_2\sin i=1.5~M_{\rm J}$ and $a=0.83$ AU
for the companion. The rms scatter of the residuals to the Keplerian
fit is 16.4 m s$^{-1}$ and the reduced $\sqrt{\chi^2}$ was 3.9 when we
adopted the measurement errors as weight of the least-squared fitting,
which may suggest existence of additional variability.
However, we found no significant periodicity in the residuals at
this stage and then we have quadratically added a jitter of
16.0 m s$^{-1}$ to the RV uncertainties to account for the variations.
The resulting Keplerian model is shown in Figure \ref{fig-HD136512}
and \ref{fig-HD136512p}  overplotted on the velocities, and its parameters
are listed in Table \ref{tbl-planets}. We did not find any significant line
shape variability for the star corresponding to the observed
RV variations (Table \ref{tbl-bisector}).

\subsection{$\nu$ Oph (HD 163917, HIP 88048)}\label{nOph}

\begin{figure}
  \begin{center}
    \FigureFile(85mm,80mm){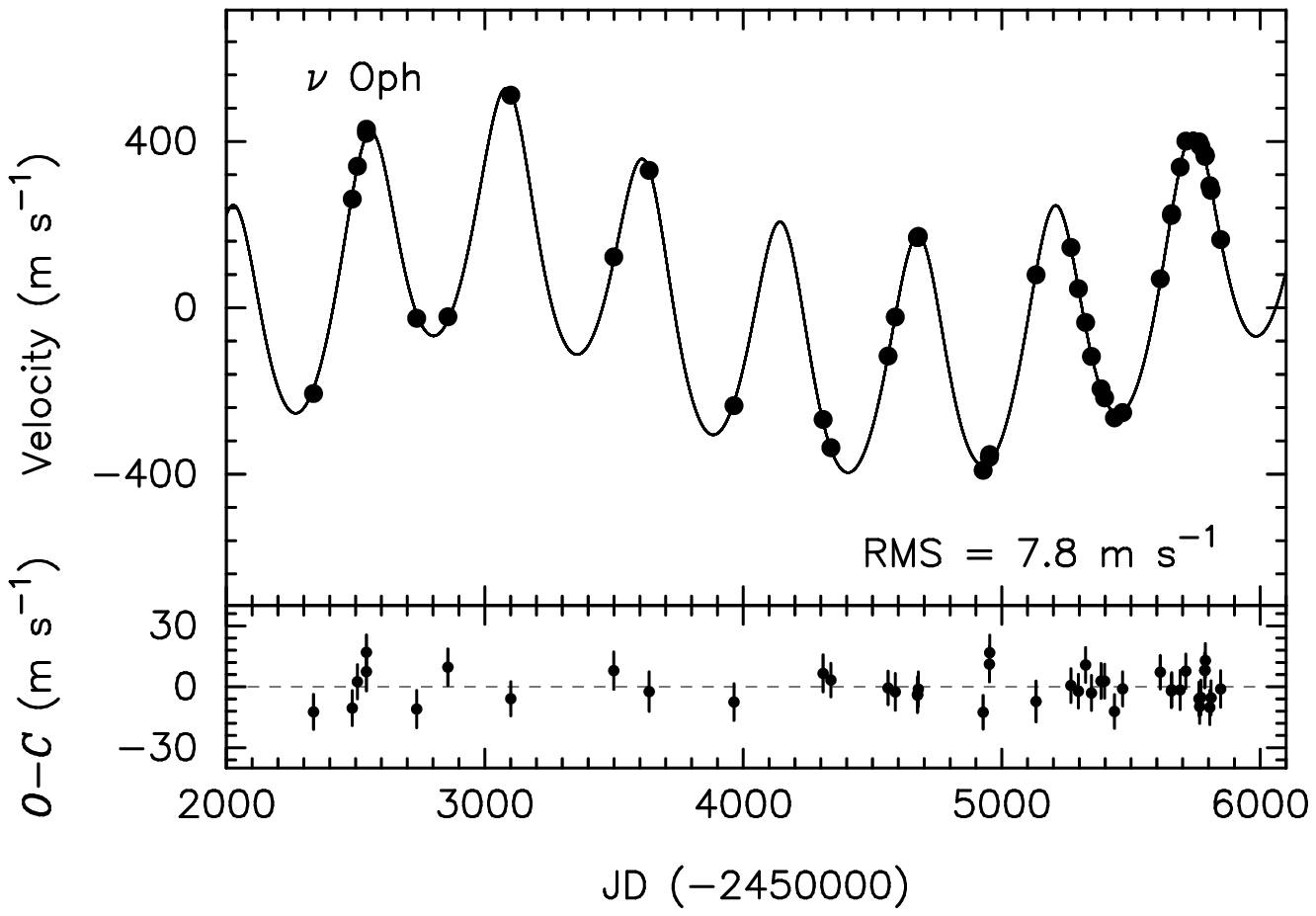}
  \end{center}
\caption{{\it Top}: Radial velocities of $\nu$ Oph observed at OAO.
The Keplerian orbit is shown by the solid line.
The error bar for each point includes the stellar jitter estimated in
section \ref{nOph}.
{\it Bottom}: Residuals to the Keplerian fit.
The rms to the fit is 7.8 m s$^{-1}$.}
\label{fig-HD163917}
\end{figure}

Brown dwarf companions to $\nu$ Oph were reported by
\citet{quirrenbach:2011} in a conference proceeding, although details of the
orbital parameters were not presetend. We collected a total of 44 RV
data of $\nu$ Oph between 2002 February and 2011 July.
The observed RVs are shown in Figure \ref{fig-HD163917}
and are listed in Table \ref{tbl-HD163917} together with their
estimated uncertainties.

Lomb-Scargle periodogram of the data exhibits a significant peak at a
period of 526 days with a $FAP<1\times10^{-5}$, and
single Keplerian model for the star yielded orbital parameters for
the companion of $P=532$ days, $K_1=331$ m s$^{-1}$,
and $e=0.15$. However, the rms scatter of the residuals to the
fit was 99.4 m s$^{-1}$ and the reduced $\sqrt{\chi^2}$ was 16.1,
which suggested the existence of additional variability. Actually
Lomb-Scagle periodogram of the residuals clearly showed a peak
with a period of $\sim$ 3000 d with a $FAP<1\times10^{-5}$.

\begin{figure}
  \begin{center}
    \FigureFile(85mm,80mm){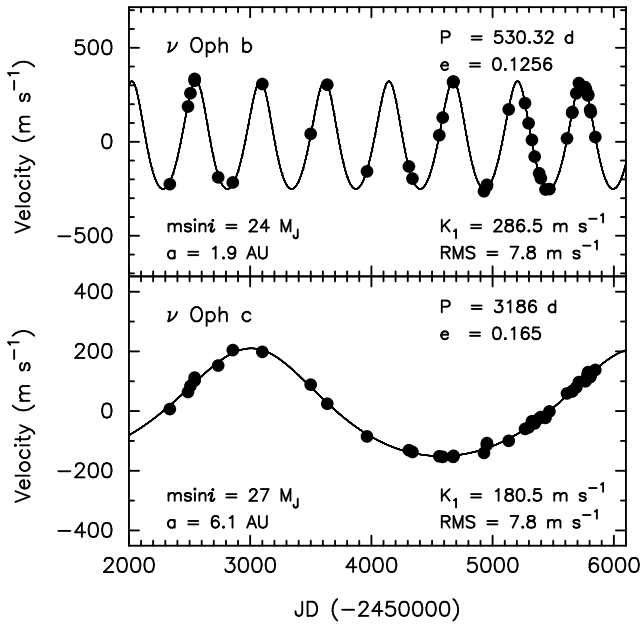}
  \end{center}
\caption{Radial velocities of $\nu$ Oph observed at OAO.
The Keplerian orbit is shown by the solid line.
The error bar for each point includes the stellar jitter estimated in
section \ref{nOph}.
{\it Top}: Inner companion with the signal from the outer companion
removed.
{\it Bottom}: Outer companion with the signal from the inner companion
removed.}
\label{fig-HD163917-each}
\end{figure}

We performed a double Keplerian fitting to the data and obtained
orbital parameters $P=530.32\pm0.35$ days, $K_1=286.5\pm1.8$ m s$^{-1}$,
and $e=0.1256\pm0.0065$ for the inner companion ($\nu$ Oph b), and
$P=3186\pm14$ days, $K_1=180.5\pm3.1$ m s$^{-1}$, and $e=0.165\pm0.013$
for the outer companion ($\nu$ Oph c).
The rms scatter of the residuals to the double Keplerian fit is 7.8 m s$^{-1}$,
and Lomb-Scargle periodogram of the residuals did not show any significant
peaks. Then to account for the variations, we have quadratically added
a jitter of 7.5 m s$^{-1}$ to the RV uncertainties.
All of the parameters are in
agreement with those presented by \citet{quirrenbach:2011}. 
Adopting a stellar mass of 3.04 $M_{\odot}$,
we obtain $m\sin i=24~M_{\rm J}$ and $a=1.9$ AU for $\nu$ Oph b,
and $m\sin i=27.0~M_{\rm J}$ and $a=6.1$ AU for $\nu$ Oph c.
The resulting Keplerian model is shown in Figure \ref{fig-HD163917}
and that for each companion is shown in Figure \ref{fig-HD163917-each}
overplotted on the velocities. The orbital parameters are listed in
Table \ref{tbl-planets}. 

For line shape analysis, we extracted templates with velocity of
$\sim300$ m s$^{-1}$ and $\sim-250$ m s$^{-1}$, and calculated
cross correlation profile between them.
As seen in Table \ref{tbl-bisector}, non-detection of any significant variations
in BVS and BVC values, and the BVD value agreeing with the difference
of the velocities between the two templates are consistent with the
orbital-motion hypothesis as the cause of the observed RV variations.

\subsection{$\kappa$ CrB (HD 142091, HIP 77655) and HD 210702 (HIP 109577)}\label{kCrB}

\begin{figure}
  \begin{center}
    \FigureFile(85mm,80mm){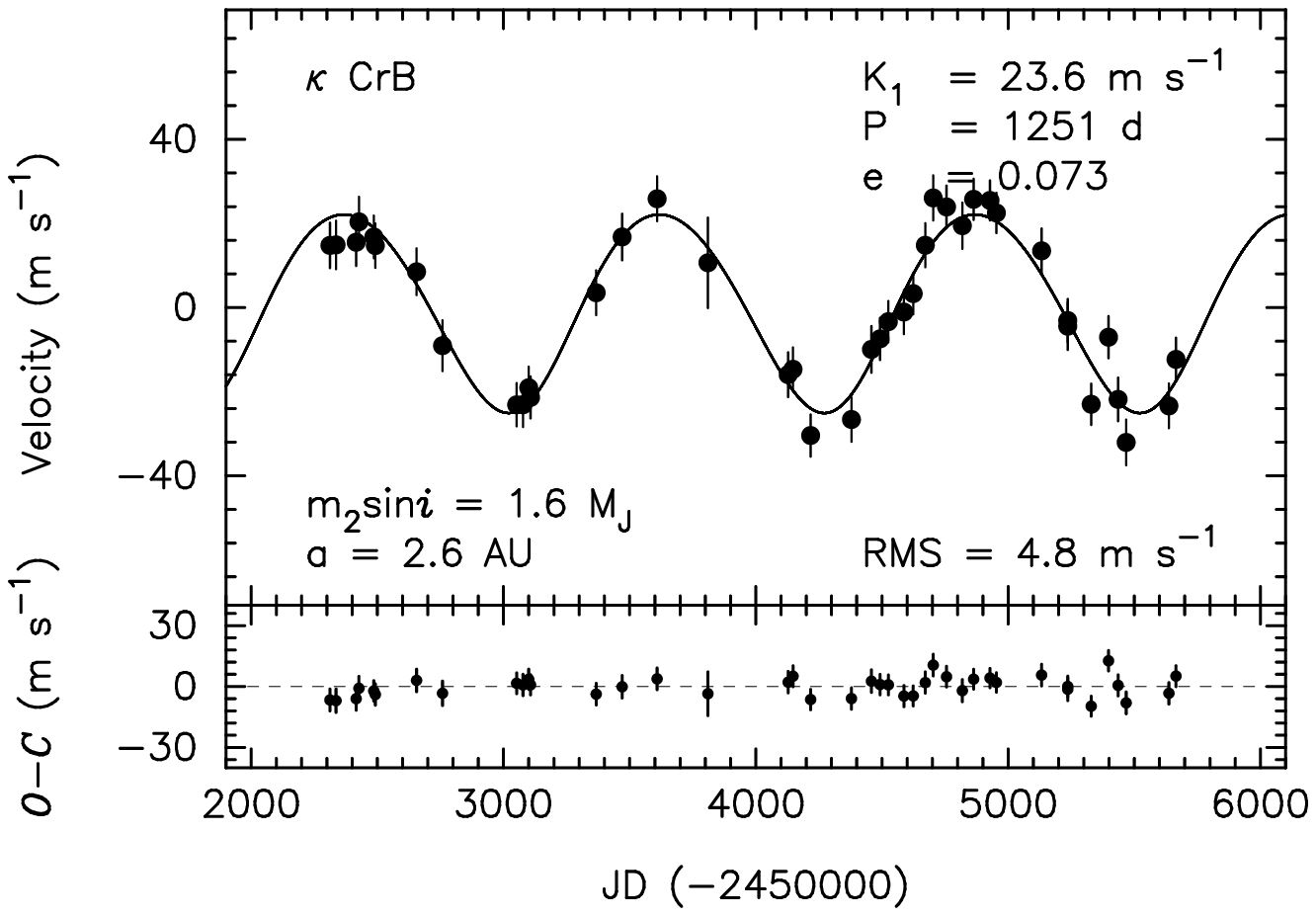}
  \end{center}
\caption{{\it Top}: Radial velocities of $\kappa$ CrB observed at OAO.
The Keplerian orbit is shown by the solid line.
The error bar for each point includes the stellar jitter estimated in
section \ref{kCrB}.
{\it Bottom}: Residuals to the Keplerian fit.
The rms to the fit is 4.8 m s$^{-1}$.}
\label{fig-HD142091}
\end{figure}

\begin{figure}
  \begin{center}
    \FigureFile(85mm,80mm){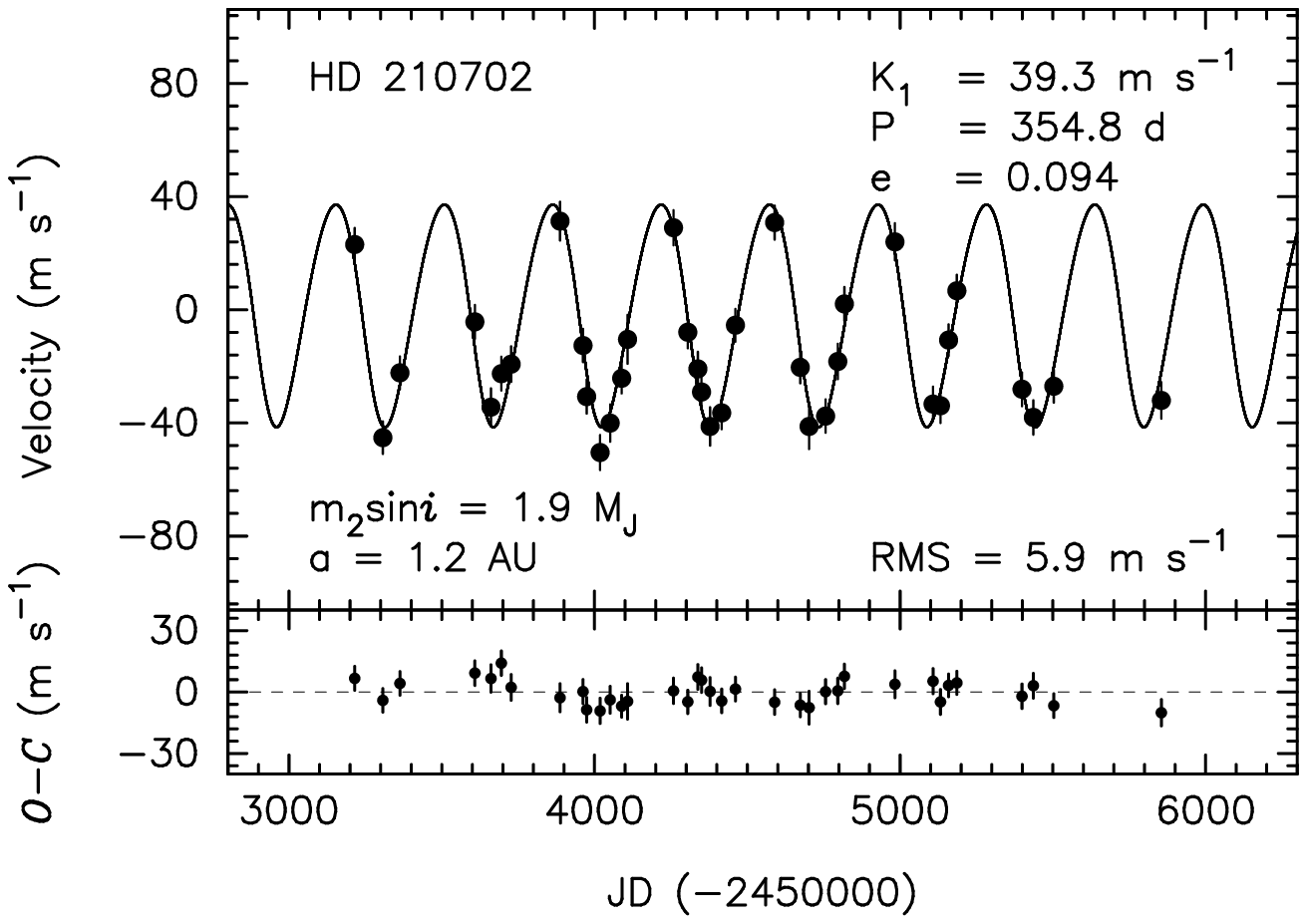}
  \end{center}
\caption{{\it Top}: Radial velocities of HD 210702 observed at OAO.
The Keplerian orbit is shown by the solid line.
The error bar for each point includes the stellar jitter estimated in
section \ref{kCrB}.
{\it Bottom}: Residuals to the Keplerian fit.
The rms to the fit is 5.9 m s$^{-1}$.}
\label{fig-HD210702}
\end{figure}

Planetary companions to $\kappa$ CrB and HD 210702 were reported
by \citet{johnson:2008} and \citet{johnson:2007a}, respectively.
We collected a total of 41 and 36 RV data of the stars between 2002
February and 2011 October, which is almost the same period of time
as those of Johnson et al.'s. The observed RVs are shown in
Figure \ref{fig-HD142091} and \ref{fig-HD210702} and are listed in
Table \ref{tbl-HD142091} and \ref{tbl-HD210702}, respectively,
together with their estimated uncertainties.
Our observed RVs for $\kappa$ CrB can be well fitted by
a single Keplerian orbit with $P=1251\pm15$ days, $K_1=23.6\pm1.1$ m s$^{-1}$,
and $e=0.073\pm0.049$. The RVs for HD 210702 can also be well fitted
by a single Keplerian orbit with $P=354.8\pm1.1$ days, $K_1=39.3\pm2.5$ m s$^{-1}$,
and $e=0.094\pm0.052$. We here adopted a stellar jitter of 4.0 m s$^{-1}$
and 4.5 m s$^{-1}$ for $\kappa$ CrB and HD 210702, respectively.
These parameters are in good agreement with
those obtained by \citet{bowler:2010}.
The resulting models are shown in Figure \ref{fig-HD142091} and
\ref{fig-HD210702}, and the parameters are listed in Table \ref{tbl-planets}.
Adopting stellar masses of 1.51 $M_{\odot}$ and 1.68 $M_{\odot}$ for
$\kappa$ CrB and HD 210702, respectively, we obtain
$m_2\sin i=1.6~M_{\rm J}$ and $a=2.6$ AU for $\kappa$ CrB b,
and $m_2\sin i=1.9~M_{\rm J}$ and $a=1.2$ AU for HD 210702 b.
We did not find any significant line shape variability for the stars
corresponding to the observed RV variations (Table \ref{tbl-bisector}).

\section{Summary and Discussion}\label{summary}

We here reported the discoveries of planetary and brown-dwarf-mass
companions to seven intermediate-mass subgiants and giants from Okayama
planet search program: four new discoveries (HD 5608 b, 75 Cet b,
$o$ UMa b, $o$ CrB b) and three independent confirmations of
previous discoveries by other groups ($\nu$ Oph bc, $\kappa$ CrB b, and
HD 210702 b). The discoveries add to the recent growing population of
substellar companions around evolved intermediate-mass stars.

HD 5608 b ($m_2\sin i=1.4~M_{\rm J}$, $a=1.9$ AU) and 75 Cet b
($m_2\sin i=3.0~M_{\rm J}$, $a=2.1$ AU)  are typical planets
orbiting intermediate-mass subgiants and giants in the point that the planets
are 1--3 $M_{\rm J}$ and resides at $a\gtrsim 1$ AU.
HD 5608 exhibits a linear trend in RV, suggesting the existence of
an outer, more massive companion. The estimated mass of the outer
companion based on equation (\ref{HD5608c}) in section \ref{HD5608} exceeds
$\sim 75~M_{\rm J}$ if it orbits at $a\gtrsim50$ AU.
75 Cet also shows possible additional periodicity of about 200 d and
1880 d in RV with velocity semiamplitude of $\sim$7--10 m s$^{-1}$.
The significance of the periodicity should be validated by collecting
more data.

\begin{figure}
  \begin{center}
    \FigureFile(85mm,80mm){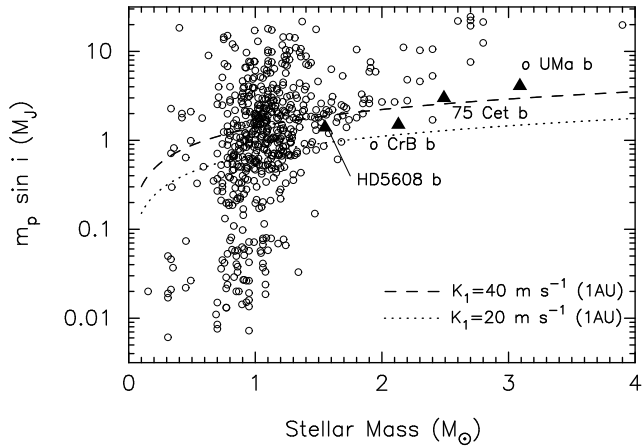}
  \end{center}
\caption{Planetary mass distribution of exoplanets detected by RV methods
against host star's mass. The data are from http://exoplanets.eu. Newly discovered planets
(HD 5608 b, 75 Cet b, $o$ UMa b, and $o$ CrB b) are labeled by filled triangles.
Dashed and dotted lines correspond to the velocity semiamplide of 40 and 20 m s$^{-1}$
for a host star, respectively, imparted by a planet at 1 AU.}
\label{fig-pmass}
\end{figure}

$o$ UMa b ($m_2\sin i=4.1~M_{\rm J}$, $a=3.9$ AU) is the first
planet candidate ($<13~M_{\rm J}$) ever discovered around stars
with $\ge3~M_{\odot}$ (see Fig. \ref{fig-pmass}). Only brown-dwarf-mass
companions had
been found around such stars. The mass of 3 $M_{\odot}$ corresponds
to that for late-B to early-A type stars on the main sequence,
which are normally rapid rotators with projected rotational velocity
$v\sin i \gtrsim 100$ km s$^{-1}$ and are also pulsating stars.
It is thus quite difficult to find planets around them by Doppler methods.
On the other hand, their evolved counter parts, GK giants, show RV
jitter of $\sim 7$ m s$^{-1}$ at most. Our discovery clearly shows that
GK giants are suitable targets to access planets around such
intermediate-mass stars.

$o$ CrB b ($m_2\sin i=1.5~M_{\rm J}$, $a=0.83$ AU) is one of the
least massive planets ever discovered around clump giants, along with
HD 100655 b ($m_2\sin i=1.7~M_{\rm J}$, $a=0.76$ AU; \cite{omiya:2011};
see Fig. \ref{fig-pmass}). 
It is generally more difficult to detect such less massive planets around
clump giants because of their larger stellar jitters compared with those of
solar-type dwarfs and subgiants. $o$ CrB shows RV semiamplitude of
32 m s$^{-1}$, which is only twice of the RMS value of the
jitter. Our discovery shows that we can detect such less massive planets
even around clump giants if we collect a large number of data points.

$\nu$ Oph hosts two brown-dwarf-mass companions, $\nu$ Oph b
($m_2\sin i=24~M_{\rm J}$) and $\nu$ Oph c ($m_2\sin i=27~M_{\rm J}$).
The ratio of the orbital period is close to 1:6 (530 d and 3186 d, respectively),
suggesting that they are in MMR.
Two such two-brown-dwarf systems have been previously reported:
one is a possible 3:2 MMR system around the clump giant BD+20 2457
(\cite{niedzielski:2009a}) and the other is around a solar-type star HD 168443
(\cite{marcy:2001}). Several scenarios are proposed for the formation of
brown-dwarf-mass companions; gravitational collapse in
protostellar clouds like stellar binary systems (\cite{bonnell:1992};
\cite{bate:2000}) and gravitational instability in cicumstellar disks
(\cite{boss:2000}; \cite{rice:2003}). Even core-accretion scenario
could form such super-massive companions with $\gtsim10M_{\rm J}$
on a certain truncation condition for gas accretion
(\cite{ida:2004}; \cite{alibert:2005}; \cite{mordasini:2007}).
The existence of such two-companions in MMR suggests that they are
formed in circumstellar disks by either gravitational instability or
core-accretion, and then experience orbital migration.
Detailed analysis of orbital stability for the system will be
presented in a separate paper.

\begin{figure}
  \begin{center}
    \FigureFile(85mm,80mm){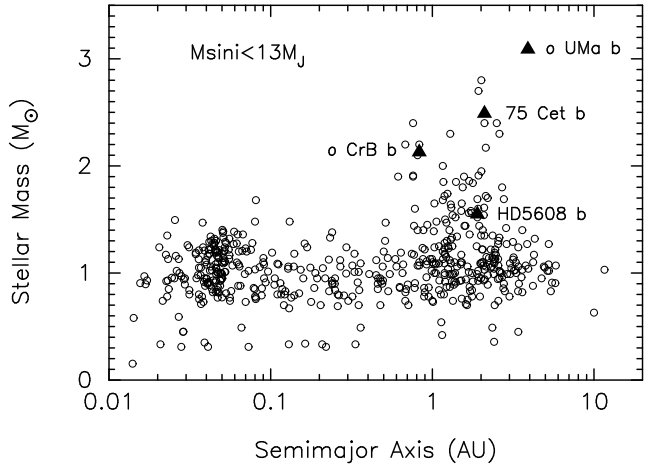}
  \end{center}
\caption{Semimajor axis distribution of exoplanets ($m_2\sin i<13M_{\rm J}$) detected
by RV methods against
host star's mass. The data are from http://exoplanets.eu. Newly discovered planets
(HD 5608 b, 75 Cet b, $o$ UMa b, and $o$ CrB b) are labeled by filled triangles.}
\label{fig-stmass}
\end{figure}

Figure \ref{fig-stmass} shows the semimajor-axis distribution of planets
detected by RV method. It has been pointed out that almost no planets
with $a\le$ 0.6 AU have been found around stars with masses
1.5--3 $M_{\odot}$, and possible lack of planetary mass companions
within 3 AU around $\sim 3~M_{\odot}$ was also pointed out (\cite{omiya:2009}).
$o$ UMa b basically follows this trend; planets tend to stay at outer
orbits as stellar mass increases.
The semimajor axis distribution should be determined by the balance
between some factors such as lifetime of a protoplanetary disk,
efficiency of planet formation, and efficiency of orbital migration.
From the numerical simulation by \citet{currie:2009}, the lack of inner
planets around intermediate-mass stars (hereafter planet desert) can be
reproduced if gas in a disk around intermediate-mass stars dissipates
more quickly than around lower mass stars, and then planets can not
migrate inward before the gas dissipation.
In the case of $\sim 3~M_{\odot}$ stars, however, the existence or
absense of the planet desert seems more sensitive to the above factors
compared with $\sim 2~M_{\odot}$ stars;
planets can form and migrate before the gas dissipation around
$\sim 3~M_{\odot}$ stars depending on assumed lifetime of the disk.
It is easier to explore inner planets around $\sim3~M_{\odot}$ giants
rather than around less massive ones because $\sim3~M_{\odot}$ giants
do not expand so largely at the tip of RGB. Then we need not
consider planet engulfment by central stars except for very
short-period ones, while planets within 0.4--1 AU around
$\sim 2~M_{\odot}$ giants could have been engulfed during RGB
phase of the central stars (\cite{kunitomo:2011}).
Planet searches around $\sim 3~M_{\odot}$ giants are thus highly encouraged.

Less massive planets ($\lesssim2M_{\rm J}$) around clump giants,
such as $o$ CrB b, should also be explored more intensively.
\citet{bowler:2010} statistically showed that relatively higher-mass
($\gtrsim2M_{\rm J}$) planets tend to exist around intermediate-mass
subgiants and excluded the same planet-mass distribution as that
for solar-type stars.
To push the detection limit from the current $K_1\sim40$ m s$^{-1}$
($\sim2~M_{\rm J}$ at $\sim1$ AU) down to $K_1\sim20$ m s$^{-1}$
($\sim1~M_{\rm J}$ at $\sim1$ AU)
for clump giants ($\sim2~M_{\odot}$)
will allow us to directly compare the mass distribution
of jovian planets for a wide range of host star's mass
(see Fig. \ref{fig-pmass}) . We are now trying to
do this by high cadence observations for a part of the sample of Okayama
Planet Search Program. The results will be presented in a forthcoming paper.
\\

This research is based on data collected at Okayama Astrophysical
Observatory (OAO), which is operated by
National Astronomical Observatory of Japan (NAOJ).
We are grateful to all the staff members of OAO for their support during
the observations. We thank students of Tokyo Institute of Technology and
Kobe University for their kind help for the observations.
BS was partly supported by MEXT's program
"Promotion of Environmental Improvement for Independence of Young
Researchers" under the Special Coordination Funds for Promoting
Science and Technology, and by Grant-in-Aid for Young Scientists (B)
20740101 from the Japan Society for the Promotion of Science (JSPS).
HI is supported by Grant-In-Aid for Scientific Research (A) 23244038
from JSPS.

This research has made use of the SIMBAD database, operated at
CDS, Strasbourg, France.


\newpage

\onecolumn
\begin{table}[h]
\rotatebox{90}{
\begin{minipage}{\textheight}
\caption{Stellar parameters}\label{tbl-stars}
\begin{center}
\begin{tabular}{cccccccc}\hline\hline
Parameter      & HD 5608 & 75 Cet & $o$ UMa & $o$ CrB & $\nu$ Oph & $\kappa$ CrB & HD 210702\\
\hline			   			   
Sp. Type         & K0 IV        & G3 III: & G4 II-III     & K0 III       & K0 III          &  K0 III-IV            & $^{\dagger}$K1 IV\\
$\pi$ (mas)     & 17.19$\pm$0.83 & 12.27$\pm$1.13 & 17.76$\pm$0.65 & 11.90$\pm$0.74 & 21.35$\pm$0.79 &  32.13$\pm$0.61 & 17.88$\pm$0.74\\
$V$                  & 5.99      & 5.36 & 3.35         & 5.51        &  3.32          &  4.79                  & 5.93\\
$B-V$              &  1.000      & 1.004 & 0.856      &  1.015     & 0.987          & 0.996               & 0.951\\ 
$A_{V}$          & 0.06         & 0.07 & 0.00         & 0.05        &  0.16           &  0.03                 & 0.05\\
$M_{V}$         & $+$2.11         & $+$0.73 &  $-$0.40  & $+$0.84  &  $-$0.19    &  $+$2.29          & $+$2.14 \\
$B.C.$            & $-$0.31   & $-$0.32 & $-$0.19   & $-$0.36   &  $-$0.28     &  $-$0.30          & $-$0.27\\
$T_{\rm eff}$ (K) & 4854$\pm$25  & 4846$\pm$18 & 5242$\pm$10 & 4749$\pm$20 &  4928$\pm$25 &  4877$\pm$25 & 4967$\pm$25\\ 
$\log g$ (cm s$^{-2}$)   & 3.03$\pm$0.08 & 2.63$\pm$0.05 & 2.64$\pm$0.03  & 2.34$\pm$0.06 &  2.63$\pm$0.09 & 3.21$\pm$0.08 & 3.19$\pm$0.08\\
$v_t$ (km s$^{-1}$)  & 1.08$\pm$0.07 & 1.26$\pm$0.08 & 1.51$\pm$0.07 & 1.39$\pm$0.06 & 1.46 $\pm$0.10 & 1.04$\pm$0.09 & 1.10$\pm$0.08\\ 
$[$Fe/H$]$  (dex)   & $+$0.06$\pm$0.04 & $+$0.00$\pm$0.04 & $-$0.09$\pm$0.02  & $-$0.29$\pm$0.03 & $+$0.13$\pm$0.05   &  $+$0.10$\pm$0.04  & $+$0.01$\pm$0.04\\
$L$ ($L_{\odot}$) &   15.1  & 53.7 & 138       & 51.2        &   123            & 12.9                 & 14.1\\
$R$ ($R_{\odot}$) &  5.5 (5.1--5.9) & 10.5 (9.5--11.5) &  14.1 (13.2--15.1) & 10.5 (9.8--11.2)  &  15.1 (14.1--16.2) & 5.0 (4.8--5.2)& 5.1 (4.8--5.5)\\
$M$ ($M_{\odot}$) & 1.55 (1.32--1.74) & 2.49 (2.22--2.51) & 3.09 (3.02--3.16)  & 2.13 (1.90--2.14) & 3.04 (2.98--3.10) & 1.51 (1.32--1.70) & 1.68 (1.50--1.84)\\
$v\sin i$ (km s$^{-1}$) & 1.37 & 1.77 & 3.83 & 2.30 & 3.21 & 1.21 & 1.99\\
$\sigma_{\rm HIP}$ (mag) & 0.008 & 0.005 & 0.004 & 0.008 & 0.004 & 0.005 & 0.007\\
\hline
\end{tabular}
\end{center}
$^{\dagger}$ The star is listed in the Hipparcos catalogue as a K1 III giant.
But judged from the position of the star on the HR diagram (Figure \ref{fig-HRD}),
the star should be better classified as a less evolved subgiant.
\\
Note -- All of the values and uncertainties for atmospheric and physical parameters
of the stars in this table are quoted by \citet{takeda:2008}. 
The uncertainties of [Fe/H], $T_{\rm eff}$, $\log g$, and $v_{\rm t}$, are internal
statistical errors and values in the parenthesis for stellar radius and mass
correspond to the range of the values assuming more realistic uncertainties in
$\Delta\log L$ corresponding to parallax errors in the Hipparcos
catalog, $\Delta\log T_{\rm eff}$ of $\pm0.01$ dex ($\sim\pm100$~K),
and $\Delta{\rm [Fe/H]}$ of $\pm0.1$ dex.
The resulting mass value may also appreciably depend on the chosen
set of theoretical evolutionary tracks (e.g., the systematic
difference as large as $\sim 0.5M_{\odot}$ for the case of
metal-poor tracks between \cite{lejeune:2001} and
\cite{girardi:2000}.; see also footnote 3 in \cite{sato:2008a}).
Please see the section 3.2  and footnote 8 of \citet{takeda:2008} for the
details of the procedure.
\end{minipage}
}
\end{table}

\begin{longtable}{ccc}
  \caption{Radial Velocities of HD 5608}\label{tbl-HD5608}
  \hline\hline
  JD & Radial Velocity & Uncertainty\\
  ($-$2450000) & (m s$^{-1}$) & (m s$^{-1}$)\\
  \hline
  \endhead
2690.92265 & 27.9 & 8.5\\
3023.94225 & $-$5.1 & 3.5\\
3216.26485 & 31.0 & 3.9\\
3363.07042 & 29.4 & 3.3\\
3405.95862 & 25.9 & 3.3\\
3426.91976 & 1.3 & 5.7\\
3579.20830 & 11.9 & 3.8\\
3612.20484 & 1.0 & 3.7\\
3729.06747 & $-$12.2 & 3.2\\
3744.06021 & $-$15.3 & 3.2\\
4018.28996 & 17.7 & 4.0\\
4054.11938 & 22.3 & 3.8\\
4089.07667 & 13.7 & 3.3\\
4338.27060 & $-$16.0 & 3.6\\
4416.06811 & $-$20.5 & 4.0\\
4461.00139 & $-$28.1 & 2.6\\
4492.98453 & $-$40.0 & 4.0\\
4756.22236 & 8.8 & 5.5\\
4796.02126 & 4.6 & 4.0\\
4818.10723 & 14.3 & 3.9\\
5101.24771 & $-$8.7 & 4.3\\
5135.26954 & $-$30.8 & 5.1\\
5162.97986 & $-$37.1 & 3.7\\
5188.09633 & $-$32.5 & 3.6\\
5205.95672 & $-$28.7 & 3.3\\
5267.91741 & $-$32.6 & 6.5\\
5351.29097 & $-$45.8 & 4.1\\
5398.26283 & $-$44.0 & 3.3\\
5438.06022 & $-$36.4 & 3.6\\
5468.19458 & $-$28.6 & 3.9\\
5499.02876 & $-$30.3 & 5.3\\
5525.05040 & $-$16.5 & 3.7\\
5550.99810 & $-$10.4 & 3.4\\
5582.91808 & 2.4 & 4.2\\
5758.20356 & $-$13.2 & 3.9\\
5770.25464 & $-$1.8 & 3.6\\
5786.19124 & $-$11.7 & 3.1\\
5790.26340 & $-$13.1 & 3.7\\
5791.27944 & $-$10.5 & 3.6\\
5811.29985 & $-$21.8 & 4.3\\
5853.19065 & $-$14.5 & 4.5\\
5879.07915 & $-$23.4 & 4.8\\
5881.13007 & $-$28.5 & 3.2\\
  \hline
\end{longtable}

\begin{longtable}{ccc}
  \caption{Radial Velocities of 75 Cet}\label{tbl-HD15779}
  \hline\hline
  JD & Radial Velocity & Uncertainty\\
  ($-$2450000) & (m s$^{-1}$) & (m s$^{-1}$)\\
  \hline
  \endhead
2311.96400 & $-$39.6 & 5.9\\
2507.25607 & 16.0 & 3.9\\
2541.26302 & 33.2 & 3.7\\
2655.94596 & 21.4 & 3.6\\
2677.01385 & 5.7 & 6.2\\
2888.23459 & $-$50.1 & 4.0\\
2896.19715 & $-$48.4 & 3.4\\
2923.20790 & $-$48.6 & 4.6\\
2974.17698 & $-$42.9 & 4.9\\
2999.96007 & $-$23.7 & 3.7\\
3023.90113 & $-$25.3 & 3.1\\
3051.92008 & $-$31.6 & 3.9\\
3214.27420 & 22.5 & 5.0\\
3246.32441 & 18.0 & 4.7\\
3284.20514 & 26.4 & 4.4\\
3306.14111 & 20.2 & 3.5\\
3311.15441 & 28.6 & 3.4\\
3332.08630 & 36.6 & 3.5\\
3339.20021 & 43.5 & 4.1\\
3362.02683 & 45.7 & 4.1\\
3366.04489 & 51.0 & 4.8\\
3401.90584 & 20.5 & 3.6\\
3424.90587 & 16.3 & 6.3\\
3424.92187 & 25.8 & 5.4\\
3447.93126 & 16.0 & 6.3\\
3579.24285 & $-$32.5 & 4.4\\
3600.23799 & $-$27.8 & 3.8\\
3635.25196 & $-$29.7 & 5.0\\
3659.26573 & $-$48.9 & 4.2\\
3693.17589 & $-$28.8 & 3.9\\
3727.08594 & $-$10.6 & 3.7\\
3743.05247 & $-$7.2 & 3.8\\
3774.92427 & 14.6 & 4.4\\
3809.91688 & 12.8 & 4.6\\
3938.29450 & 49.2 & 4.4\\
3963.29661 & 34.6 & 3.8\\
4018.17496 & 38.3 & 3.8\\
4051.06681 & 15.6 & 3.5\\
4087.15142 & 6.0 & 5.5\\
4089.13018 & 14.5 & 4.1\\
4112.96188 & $-$2.3 & 4.2\\
4142.97154 & $-$1.0 & 3.4\\
4170.91468 & $-$22.9 & 5.2\\
4305.26627 & $-$43.5 & 4.7\\
4338.25207 & $-$49.6 & 4.5\\
4378.20983 & $-$39.0 & 4.2\\
4416.05343 & $-$30.2 & 3.7\\
4458.10607 & $-$31.9 & 4.4\\
4493.00758 & 0.0 & 3.8\\
4526.92907 & 17.8 & 4.6\\
4702.30722 & 30.5 & 3.4\\
4756.19459 & 30.6 & 3.8\\
4796.06006 & 20.8 & 3.7\\
4818.04972 & $-$1.9 & 4.0\\
4856.01989 & $-$14.6 & 3.7\\
5075.31961 & $-$30.4 & 4.0\\
5131.15632 & $-$11.6 & 3.2\\
5183.04922 & 5.1 & 3.8\\
5234.98763 & 20.8 & 3.5\\
5266.91271 & 39.1 & 3.8\\
5398.29099 & 43.8 & 3.9\\
5439.25969 & 24.8 & 3.5\\
5468.17983 & 21.2 & 3.9\\
5502.14696 & 27.9 & 4.0\\
5525.07454 & 13.2 & 4.2\\
5525.20750 & 20.9 & 4.2\\
5545.09659 & $-$2.7 & 3.9\\
5581.98008 & 3.5 & 3.6\\
5759.28687 & $-$16.7 & 3.8\\
5786.29592 & $-$9.4 & 3.9\\
5811.27161 & $-$18.8 & 3.7\\
5854.18481 & $-$4.7 & 3.8\\
5879.12898 & $-$5.0 & 4.1\\
5920.00604 & 22.0 & 4.2\\
  \hline
\end{longtable}

\begin{longtable}{ccc}
  \caption{Radial Velocities of $o$ UMa}\label{tbl-HD71369}
  \hline\hline
  JD & Radial Velocity & Uncertainty\\
  ($-$2450000) & (m s$^{-1}$) & (m s$^{-1}$)\\
  \hline
  \endhead
3000.21377 & 12.9 & 4.6\\
3081.12751 & 33.2 & 4.8\\
3311.25752 & 28.0 & 4.4\\
3427.12651 & 22.6 & 5.1\\
3500.01480 & $-$9.6 & 5.5\\
3660.31478 & $-$17.7 & 4.5\\
3729.21876 & $-$29.6 & 4.7\\
3812.09846 & $-$20.4 & 4.6\\
3886.96433 & $-$17.7 & 6.5\\
4022.32303 & $-$47.9 & 6.2\\
4123.15218 & $-$19.6 & 5.2\\
4186.04763 & $-$14.6 & 4.6\\
4216.06204 & $-$17.5 & 5.0\\
4461.33804 & 16.6 & 4.1\\
4525.12025 & 14.7 & 4.6\\
4587.94612 & 28.3 & 6.3\\
4796.33191 & 37.5 & 4.6\\
4887.99762 & 26.9 & 5.4\\
4925.03379 & 43.4 & 4.8\\
5166.10163 & 0.1 & 5.1\\
5205.19728 & $-$7.5 & 4.5\\
5233.13003 & $-$7.5 & 5.1\\
5346.97821 & $-$26.8 & 5.0\\
5502.24894 & $-$39.8 & 6.2\\
5558.23852 & $-$27.5 & 4.7\\
5625.05259 & $-$31.8 & 4.5\\
  \hline
\end{longtable}

\begin{longtable}{ccc}
  \caption{Radial Velocities of $o$ CrB}\label{tbl-HD136512}
  \hline\hline
  JD & Radial Velocity & Uncertainty\\
  ($-$2450000) & (m s$^{-1}$) & (m s$^{-1}$)\\
  \hline
  \endhead
2350.23364 & 48.9 & 6.7\\
2487.03829 & $-$18.3 & 4.3\\
2507.02176 & $-$12.1 & 4.0\\
2538.95959 & 31.3 & 8.7\\
2541.92576 & 34.9 & 3.9\\
2736.22496 & 32.2 & 6.2\\
2756.22924 & 24.6 & 8.6\\
2858.04958 & $-$5.7 & 3.6\\
2991.38250 & $-$38.0 & 4.8\\
3002.31699 & $-$49.9 & 4.7\\
3009.35443 & $-$33.0 & 4.4\\
3024.30929 & $-$39.3 & 4.2\\
3030.33746 & $-$10.6 & 4.6\\
3052.21246 & 12.0 & 4.6\\
3077.21046 & 11.1 & 4.2\\
3080.24297 & 4.4 & 4.6\\
3100.14408 & 27.0 & 3.9\\
3106.13778 & 19.9 & 8.1\\
3113.09319 & $-$2.0 & 4.6\\
3123.13165 & 21.9 & 6.9\\
3123.15778 & 33.8 & 4.5\\
3131.12074 & 47.0 & 3.6\\
3154.07703 & 24.0 & 4.4\\
3201.02789 & $-$4.7 & 3.4\\
3213.05306 & $-$3.5 & 8.2\\
3213.06729 & $-$10.6 & 9.2\\
3215.05894 & $-$14.4 & 3.9\\
3231.05763 & $-$12.6 & 4.1\\
3245.98782 & $-$7.4 & 3.7\\
3248.99489 & $-$5.0 & 5.2\\
3289.89738 & 43.9 & 3.8\\
3306.87783 & 66.8 & 3.7\\
3310.88572 & 46.9 & 8.0\\
3332.36721 & 1.5 & 5.0\\
3335.35492 & $-$0.3 & 5.4\\
3339.37537 & $-$1.2 & 4.2\\
3365.33383 & $-$40.9 & 4.1\\
3366.33105 & $-$26.3 & 4.6\\
3403.28149 & $-$1.6 & 3.9\\
3424.19652 & $-$25.6 & 5.1\\
3447.27715 & $-$3.8 & 5.4\\
3468.13912 & 2.4 & 5.9\\
3495.12788 & 17.3 & 4.9\\
3580.01261 & $-$17.7 & 5.3\\
3655.89792 & 17.3 & 3.5\\
3744.34486 & $-$45.1 & 3.5\\
3775.29240 & $-$26.0 & 3.9\\
3814.25797 & 5.1 & 3.8\\
3854.14773 & 6.8 & 7.1\\
3889.13885 & 22.8 & 5.7\\
3962.08374 & $-$34.1 & 4.8\\
4018.91296 & $-$14.7 & 5.9\\
4089.34917 & 7.7 & 3.8\\
4127.30504 & $-$51.9 & 4.6\\
4146.27743 & $-$50.1 & 5.7\\
4169.13495 & $-$44.0 & 4.7\\
4195.24643 & 1.0 & 3.9\\
4216.18929 & 42.0 & 3.3\\
4253.06244 & 10.9 & 6.9\\
4305.03318 & $-$49.2 & 3.7\\
4337.99685 & $-$15.0 & 4.1\\
4378.91908 & $-$16.9 & 4.3\\
4458.35903 & 14.4 & 4.2\\
4492.34343 & 5.0 & 5.0\\
4524.27230 & $-$29.8 & 3.9\\
4554.21774 & 11.7 & 4.4\\
4587.19292 & 37.8 & 4.3\\
4623.01424 & 56.0 & 3.6\\
4672.00836 & $-$0.0 & 3.8\\
4702.95761 & $-$15.6 & 4.0\\
4757.89315 & $-$0.7 & 4.2\\
4818.34183 & 11.4 & 5.6\\
4864.31216 & $-$23.2 & 3.9\\
4928.13244 & $-$17.8 & 3.5\\
4953.09860 & 53.8 & 4.0\\
5182.37706 & 52.0 & 3.6\\
5270.27221 & $-$44.0 & 5.5\\
5329.27285 & $-$2.7 & 3.6\\
5400.04996 & 7.3 & 3.7\\
5439.96902 & $-$2.6 & 3.6\\
5626.27038 & $-$56.9 & 3.7\\
5657.24529 & $-$21.4 & 4.1\\
5717.11504 & 23.0 & 4.3\\
5785.95886 & 10.3 & 6.5\\
5853.91399 & $-$9.7 & 4.4\\
  \hline
\end{longtable}

\begin{longtable}{ccc}
  \caption{Radial Velocities of $\nu$ Oph}\label{tbl-HD163917}
  \hline\hline
  JD & Radial Velocity & Uncertainty\\
  ($-$2450000) & (m s$^{-1}$) & (m s$^{-1}$)\\
  \hline
  \endhead
2337.36482 & $-$206.2 & 4.3\\
2487.10137 & 261.5 & 4.2\\
2507.07715 & 340.3 & 3.8\\
2541.95527 & 419.9 & 5.8\\
2541.95830 & 429.4 & 4.0\\
2736.30695 & $-$25.2 & 5.3\\
2857.03714 & $-$21.5 & 5.0\\
3100.27680 & 511.3 & 3.7\\
3499.17826 & 122.5 & 5.3\\
3636.02517 & 330.5 & 6.2\\
3963.95123 & $-$235.1 & 4.9\\
4309.11301 & $-$268.7 & 5.2\\
4338.99473 & $-$336.4 & 3.4\\
4560.23968 & $-$116.0 & 3.5\\
4588.26944 & $-$22.3 & 5.0\\
4674.02304 & 169.2 & 4.1\\
4674.02538 & 168.7 & 4.1\\
4677.06734 & 171.7 & 3.8\\
4928.33015 & $-$390.8 & 3.5\\
4952.30133 & $-$359.5 & 4.5\\
4953.31022 & $-$353.2 & 4.1\\
5132.93025 & 79.4 & 6.7\\
5267.33386 & 145.0 & 3.5\\
5296.32995 & 45.8 & 3.5\\
5324.28497 & $-$35.2 & 4.0\\
5347.15367 & $-$117.0 & 4.1\\
5384.18718 & $-$194.7 & 4.2\\
5398.04796 & $-$216.6 & 3.7\\
5436.02615 & $-$264.5 & 3.5\\
5467.92821 & $-$251.8 & 3.7\\
5613.38119 & 69.8 & 3.5\\
5656.34153 & 222.2 & 3.3\\
5657.34080 & 226.5 & 3.8\\
5690.21013 & 338.5 & 6.0\\
5712.15981 & 401.0 & 3.8\\
5763.07019 & 399.3 & 3.4\\
5766.07583 & 390.9 & 3.6\\
5770.06689 & 388.6 & 4.0\\
5784.95022 & 369.3 & 4.1\\
5787.00550 & 364.2 & 3.6\\
5788.00217 & 366.3 & 3.7\\
5804.92863 & 293.8 & 3.8\\
5809.97268 & 282.5 & 3.7\\
5846.91912 & 164.1 & 4.9\\  
  \hline
\end{longtable}

\begin{longtable}{ccc}
  \caption{Radial Velocities of $\kappa$ CrB}\label{tbl-HD142091}
  \hline\hline
  JD & Radial Velocity & Uncertainty\\
  ($-$2450000) & (m s$^{-1}$) & (m s$^{-1}$)\\
  \hline
  \endhead
2312.26533 & 14.8 & 3.6\\
2336.33946 & 14.9 & 4.1\\
2416.12452 & 15.5 & 4.0\\
2427.20074 & 20.4 & 4.4\\
2486.03119 & 16.8 & 3.1\\
2492.10142 & 14.7 & 3.4\\
2655.39085 & 8.5 & 3.9\\
2758.18803 & $-$9.0 & 4.5\\
3052.22450 & $-$23.1 & 3.2\\
3077.22294 & $-$23.1 & 3.4\\
3100.19006 & $-$19.1 & 3.0\\
3107.22020 & $-$21.4 & 3.0\\
3367.32270 & 3.5 & 3.5\\
3470.23230 & 16.8 & 3.8\\
3609.05756 & 25.9 & 3.5\\
3810.20587 & 10.6 & 10.0\\
4128.36571 & $-$16.0 & 3.5\\
4147.37825 & $-$14.7 & 3.3\\
4217.18189 & $-$30.4 & 3.0\\
4378.93344 & $-$26.6 & 3.4\\
4458.36899 & $-$9.9 & 3.8\\
4492.35201 & $-$7.4 & 3.0\\
4525.33969 & $-$3.4 & 2.8\\
4587.20318 & $-$1.1 & 3.4\\
4624.06530 & 3.3 & 2.9\\
4672.01408 & 14.8 & 3.4\\
4702.96763 & 26.1 & 3.5\\
4755.90487 & 23.9 & 3.2\\
4818.37719 & 19.5 & 3.7\\
4863.38001 & 25.7 & 2.8\\
4928.14456 & 25.5 & 2.6\\
4953.14259 & 22.5 & 2.6\\
5132.89637 & 13.5 & 3.5\\
5236.30554 & $-$4.4 & 4.0\\
5236.31346 & $-$3.1 & 3.2\\
5329.28650 & $-$23.0 & 2.8\\
5398.07959 & $-$7.0 & 3.0\\
5435.97751 & $-$21.8 & 3.3\\
5467.91961 & $-$32.1 & 3.6\\
5637.20790 & $-$23.4 & 3.5\\
5665.15483 & $-$12.3 & 3.2\\
  \hline
\end{longtable}

\begin{longtable}{ccc}
  \caption{Radial Velocities of HD 210702}\label{tbl-HD210702}
  \hline\hline
  JD & Radial Velocity & Uncertainty\\
  ($-$2450000) & (m s$^{-1}$) & (m s$^{-1}$)\\
  \hline
  \endhead
3215.08376 & 23.0 & 3.7\\
3306.99833 & $-$45.2 & 3.6\\
3362.91310 & $-$22.3 & 3.6\\
3608.23193 & $-$4.3 & 3.9\\
3661.15055 & $-$34.4 & 4.9\\
3695.02464 & $-$22.6 & 3.9\\
3726.91613 & $-$19.2 & 4.4\\
3887.29767 & 31.3 & 5.2\\
3962.22805 & $-$12.6 & 3.8\\
3974.12967 & $-$30.7 & 4.0\\
4018.05863 & $-$50.4 & 4.2\\
4051.04081 & $-$40.1 & 4.8\\
4088.91609 & $-$24.2 & 2.9\\
4107.92459 & $-$10.5 & 7.4\\
4258.28341 & 29.0 & 4.2\\
4305.16396 & $-$7.9 & 3.5\\
4338.13598 & $-$20.9 & 4.1\\
4350.11134 & $-$29.1 & 3.9\\
4378.17418 & $-$41.3 & 5.1\\
4416.00348 & $-$36.4 & 3.6\\
4460.96815 & $-$5.5 & 3.6\\
4589.29655 & 30.8 & 4.0\\
4673.25542 & $-$20.3 & 3.5\\
4702.27411 & $-$41.2 & 6.6\\
4756.09323 & $-$37.6 & 3.8\\
4795.97776 & $-$18.3 & 4.3\\
4817.97314 & 2.1 & 3.9\\
4983.27072 & 24.0 & 4.8\\
5108.06283 & $-$33.3 & 4.1\\
5132.05624 & $-$33.9 & 4.1\\
5159.05429 & $-$10.7 & 3.2\\
5185.87590 & 6.7 & 3.4\\
5399.16553 & $-$28.1 & 3.9\\
5436.27068 & $-$38.1 & 4.1\\
5503.00859 & $-$27.0 & 3.7\\
5855.07749 & $-$32.1 & 4.6\\
  \hline
\end{longtable}

\begin{table}
\rotatebox{90}{
\begin{minipage}{\textheight}
  \caption{Orbital Parameters}\label{tbl-planets}
  \begin{center}
    \begin{tabular}{lrrrrrrrr}
  \hline\hline
  Parameter      & HD 5608 b & 75 Cet b & $o$ UMa b & $o$ CrB b & $\nu$ Oph b & $\nu$ Oph c & $\kappa$ CrB b & HD 210702 b\\
  \hline
$P$ (days)                    & 792.6$\pm$7.7 & 691.9$\pm$3.6 & 1630$\pm$35  & 187.83$\pm$0.54  & 530.32$\pm$0.35 & 3186$\pm$14 & 1251$\pm$15 & 354.8$\pm$1.1\\
$K_1$ (m s$^{-1}$)     & 23.5$\pm$1.6  & 38.3$\pm$2.0 & 33.6$\pm$2.1       & 32.25$\pm$2.8    & 286.5$\pm$1.8 & 180.5$\pm$3.1 & 23.6$\pm$1.1 & 39.3$\pm$2.5\\
$e$                                 & 0.190$\pm$0.061 & 0.117$\pm$0.048 & 0.130$\pm$0.065     & 0.191$\pm$0.085 & 0.1256$\pm$0.0065 & 0.165$\pm$0.013 & 0.073$\pm$0.049 & 0.094$\pm$0.052\\
$\omega$ (deg)           & 269$\pm$22   & 165$\pm$28 & 58$\pm$42       & 79$\pm$36 & 9.6$\pm$3.8 & 4.6$\pm$4.5 & 210$\pm$51 & 126$\pm$47\\
$T_p$    (JD$-$2,450,000)     & 2327$\pm$61 & 2213$\pm$54 & 3400$\pm$170  & 2211$\pm$17 & 2034.2$\pm$6.6 & 3038$\pm$38 & 1860$\pm$180 & 2205$\pm$45\\
$\dot{\gamma}$ (m s$^ {-1}$ yr$^{-1}$)  & $-$5.51$\pm$0.45 & 0 (fixed) & 0 (fixed) & 0 (fixed) & 0 (fixed) & 0 (fixed) & 0 (fixed) & 0 (fixed)\\
$a_1\sin i$ (10$^{-3}$AU)     & 1.68$\pm$0.11 & 2.42$\pm$0.13 & 5.00$\pm$0.33 & 0.547$\pm$0.043 & 13.853$\pm$0.090 & 52.13$\pm$0.85 & 2.71$\pm$0.14 & 1.277$\pm$0.085\\
$f_1(m)$ (10$^{-7}M_{\odot}$) & 0.0101$\pm$0.0020 & 0.0395$\pm$0.0062 & 0.063$\pm$0.012 & 0.0062$\pm$0.0016 & 12.61$\pm$0.25 & 18.62$\pm$0.89 & 0.0169$\pm$0.0025 & 0.0221$\pm$0.0046\\
$m_2\sin i$ ($M_{\rm J}$)     & 1.4  & 3.0 & 4.1 & 1.5  & 24 & 27 & 1.6 & 1.9\\
$a$ (AU)                       & 1.9   & 2.1  & 3.9    & 0.83  & 1.9   & 6.1 & 2.6 & 1.2\\
$N_{\rm obs}$              & 43   & 74  & 26     & 85      & 44   &  44  & 41 & 36\\
RMS (m s$^{-1}$)           & 6.3  & 10.8  & 7.6   & 16.4   & 7.8  &  7.8 & 4.8 & 5.9\\
$^{\dagger}$Reduced $\sqrt{\chi^2}$     & 1.7 & 2.7  & 1.6 & 3.9  & 2.2  & 2.2 & 1.6 & 1.6\\
$\sigma_{\rm jitter}$ (m s$^{-1}$) & 5.0 & 10 & 6.5 & 16 & 7.5 & 7.5 & 4.0 & 4.5\\
  \hline
    \end{tabular}
  \end{center}
$^{\dagger}$ The value without taking account of stellar jitter $\sigma_{\rm jitter}$
\end{minipage}
}
\end{table}

\begin{table}
\rotatebox{90}{
\begin{minipage}{\textheight}
  \caption{Bisector Quantities}\label{tbl-bisector}
  \begin{center}
    \begin{tabular}{lrrrrrrr}
  \hline\hline
  Bisector Quantities & HD 5608 & 75 Cet & $o$ UMa & $o$ CrB & $\nu$ Oph & $\kappa$ CrB & HD 210702\\
  \hline
Bisector Velocity Span (BVS) (m s$^{-1}$) & 6.0$\pm$4.7 & $-$5.8$\pm$4.3 & $-$3.3$\pm$5.5 & 5.6$\pm$6.5 & 7.2$\pm$5.7 & $-$4.7$\pm$3.9 & $-$8.2$\pm$6.7\\
Bisector Velocity Curve (BVC) (m s$^{-1}$) & 3.1$\pm$3.4 & 3.4$\pm$2.2 & 0.4$\pm$3.8 & 0.4$\pm$3.6 & $-$5.9$\pm$3.4 & 3.9$\pm$2.5 & $-$1.4$\pm$4.5\\
Bisector Velocity Displacement (BVD) (m s$^{-1}$) & $-$41$\pm$11 & $-$71$\pm$9 & $-$40$\pm$11 & $-$77$\pm$12 & $-$570$\pm$10 & $-$31$\pm$7 & $-$37$\pm$13\\
  \hline
    \end{tabular}
  \end{center}
\end{minipage}
}
\end{table}


\begin{thebibliography}{}
\bibitem[Alibert et al.(2005)]{alibert:2005} Alibert, Y., Mordasini, C.,
    Benz, W., Winisdoerffer, C. 2005, \aap, 434, 343
\bibitem[Arenou et al.(1992)]{arenou:1992} Arenou, F., Grenon, M., \&
    Gomez, A. 1992, \aap, 258, 104
\bibitem[Bate(2000)]{bate:2000} Bate, M. R.  2000,
    MNRAS, 314, 33
\bibitem[Bonnell \& Bastien(1992)]{bonnell:1992} Bonnell, I. \& Bastien, P. 1992,
    ApJ, 401, 654
\bibitem[Boss(2000)]{boss:2000} Boss, A. P.  2000,
    ApJ, 536, L101
\bibitem[Bowler et al.(2010)]{bowler:2010} Bowler, B.P., et al., \apj, 710, 1365
\bibitem[Butler et al.(1996)]{butler:1996} Butler, R. P., Marcy, G. W.,
    Williams, E., McCarthy, C., Dosanjh, P., \& Vogt, S. S.  1996,
    \pasp, 108, 500
\bibitem[Cumming et al.(1999)]{cumming:1999} Cumming, A., Marcy, G.~W., \&
Butler, R.~P., 1999, \apj, 526, 890
\bibitem[Currie(2009)]{currie:2009} Currie, T. 2009, \apj, 694, 171
\bibitem[de Medeiros et al.(2009)]{demedeiros:2009} de Medeiros, J.R., et al.,
\aap, 504, 617
\bibitem[D$\ddot{\rm{o}}$llinger et al.(2009)]{dollinger:2009}
D$\ddot{\rm{o}}$llinger, M.P., Hatzes, A.P.,
    Pasquini, L., Guenther, E.W., \& Hartmann, M. 2009, \aap, 505, 1311
\bibitem[ESA(1997)]{esa:1997} ESA.  1997,
    The {\it Hipparcos} and Tycho Catalogues (ESA SP-1200; Noordwijk: ESA)
    \aap, 394, 5
\bibitem[Frink et al.(2002)]{frink:2002} Frink, S., Mitchell, D.S.,
    Quirrenbach, A., Fischer, D., Marcy, G.W., \& Butler, R.P. 2002,
    \apj, 576, 478
\bibitem[Girardi et al.(2000)]{girardi:2000} Girardi, L., Bressan, A.,
    Bertelli, G., \& Chiosi, C. 2000, \aaps, 141, 371
\bibitem[Hatzes et al.(2005)]{hatzes:2005} Hatzes, A.P., Guenther, E.W.,
    Endl, M., Cochran, W.D., D$\ddot{\rm{o}}$llinger, M.P., \& Bedalov, A.
    2005, \aap, 437, 743
\bibitem[Hatzes et al.(2006)]{hatzes:2006} Hatzes, A.P., et al.
    2006, \aap, 457, 335
\bibitem[Ida \& Lin (2004)]{ida:2004} Ida, S. \& Lin, D.N.C.
    2004, \apj, 616, 567
\bibitem[Izumiura(1999)]{izumiura:1999} Izumiura, H. 1999,
    in Proc. 4th East Asian Meeting on Astronomy, ed. P.S. Chen
    (Kunming: Yunnan Observatory), 77
\bibitem[Johnson et al.(2007a)]{johnson:2007a} Johnson, J.A., et al. 2007a,
    \apj, 665, 785
\bibitem[Johnson et al.(2007b)]{johnson:2007b} Johnson, J.A., et al. 2007b,
    \apj, 670, 833
\bibitem[Johnson et al.(2008)]{johnson:2008} Johnson, J.A., et al. 2008,
    \apj, 675, 784
\bibitem[Johnson et al.(2010)]{johnson:2010} Johnson, J.A., et al. 2010,
    \pasp, 122, 905
\bibitem[Johnson et al.(2011a)]{johnson:2011a} Johnson, J.A., et al. 2011,
    \apjs, 197, 26
\bibitem[Johnson et al.(2011b)]{johnson:2011b} Johnson, J.A., et al. 2011,
    \aj, 141, 16
\bibitem[Kambe et al.(2002)]{kambe:2002} Kambe, E., et al. 2002,
    \pasj, 54, 865
\bibitem[Kunitomo et al.(2011)]{kunitomo:2011} Kunitomo, M., Ikoma, M.,
Sato, B., Katsuta, Y., \& Ida, S. 2011, \apj, 737, 66
\bibitem[Kurucz(1993)]{kurucz:1993} Kurucz, R. L. 1993, Kurucz CD-ROM,
    No. 13 (Harvard-Smithsonian Center for Astrophysics)
\bibitem[Liu et al.(2008)]{liu:2008} Liu, Y.-J., et al. 2008, \apj, 672, 553 
\bibitem[Liu et al.(2009)]{liu:2009} Liu, Y.-J., Sato, B., Zhao, G., \& Ando, H.
    2009, RAA, 9, L1 
\bibitem[Lejeune \& Schaerer(2001)]{lejeune:2001} Lejeune, T., \& Schaerer, D.
    2001, \aap, 366, 538
\bibitem[Lovis \& Mayor(2007)]{lovis:2007} Lovis, C., \& Mayor, M. 2007,
    \aap, 472, 657
\bibitem[Marcy et al.(2001)]{marcy:2001} Marcy, G.W., et al., 2001, \apj, 555, 418
\bibitem[Marois et al.(2008)]{marois:2008} Marois, C.. et al. 2008, Science, 322, 1348
\bibitem[Mordasini et al.(2007)]{mordasini:2007} Mordasini, C.,
    Alibert, Y., Benz, W., \& Naef, D. 2007, in ``Extreme Solar Systems'',
    ASP Conf. Ser. Vol. 398, 235
\bibitem[Niedzielski et al.(2009a)]{niedzielski:2009a} Niedzielski, A., Gozdziewski, K.,
    Wolszczan, A., Konacki, M., Nowak, G., \& Zielinski, P. 2009a, \apj, 693, 276
\bibitem[Niedzielski et al.(2009b)]{niedzielski:2009b} Niedzielski, A., Nowak, G.,
    Adamow, M., \& Wolszczan, A. 2009b, \apj, 707, 768
\bibitem[Omiya et al.(2009)]{omiya:2009} Omiya, M., et al., 2009, \pasj, 61, 825
\bibitem[Omiya et al.(2011)]{omiya:2011} Omiya, M., et al., 2011, \pasj, in press
   (arXiv:1111.3746v1)
\bibitem[Pasquini et al.(2007)]{pasquini:2007} Pasquini, L., et al. 2007,
    \aap, 473, 979
\bibitem[Press et al.(1989)]{press:1989} Press, W.H., Flannery, B.P., Teukolsky, S. A.,
    Vetterling, W. T. 1989, "Numerical Recipe in C. The art of scientific computing",
    Cambridge Univ. Press
\bibitem[Quirrenbach et al.(2011)]{quirrenbach:2011} Quirrenbach, A., Reffert, S., \&
    Bergmann, C. 2011, AIPC, 1331, 102
\bibitem[Rice et al.(2003)]{rice:2003} Rice, W.K.M., Armitage, P.J.,
    Bonnell, I.A., Bate, M.R., Jeffers, S.V., \& Vine, S.G., 2003,
    MNRAS, 346, L36
\bibitem[Sato et al.(2002)]{sato:2002} Sato, B., Kambe, E.,
    Takeda, Y., Izumiura, H., \& Ando, H.  2002,
    \pasj, 54, 873
\bibitem[Sato et al.(2003)]{sato:2003} Sato, B., et al. 2003, \apj,
    597, L157
\bibitem[Sato et al.(2005)]{sato:2005} Sato, B., Kambe, E.,
    Takeda, Y., Izumiura, H., Masuda, S., \& Ando, H.  2005,
    \pasj, 57, 97
\bibitem[Sato et al.(2007)]{sato:2007} Sato, B., et al. 2007,
    \apj, 661, 527
\bibitem[Sato et al.(2008a)]{sato:2008a} Sato, B., et al. 2008a,
    \pasj, 60, 539
\bibitem[Sato et al.(2008b)]{sato:2008b} Sato, B., et al. 2008b,
    \pasj, 60, 1317
\bibitem[Sato et al.(2010)]{sato:2010} Sato, B., et al. 2010,
    \pasj, 62, 1063
\bibitem[Scargle(1982)]{scargle:1982} Scargle, J. D. 1982,
    \apj, 263, 835
\bibitem[Setiawan et al.(2005)]{setiawan:2005} Setiawan, J., et al. 2005,
    \aap, 437, 31
\bibitem[Takeda et al.(2002)]{takeda:2002} Takeda, Y., Ohkubo, M., \&
    Sadakane, K. 2002, \pasj, 54, 451
\bibitem[Takeda et al.(2008)]{takeda:2008} Takeda, Y., Sato, B., \& Murata, D.,
    2008, \pasj, 60, 781
\bibitem[Wang et al.(2011)]{wang:2011} Wang, L., et al., 2011, RAA,
    in press (arXiv:1110.0559v1)
\bibitem[Winn et al.(2009)]{winn:2009} Winn, J. N., Johnson, J.A., Albrecht, S.,
    Howard, A.W., Marcy, G.W., Crossfield, I.J., \& Holman, M.J., 2009, \apjl,
     703, L99
\bibitem[Wittenmyer et al.(2011)]{wittenmyer:2011} Wittenmyer, R. A., Endl, M.,
   Wang, L., Johnson, J. A., Tinney, C. G., \& O'Toole, S. J. 2011, \apj,
   743, 184
\end{thebibliography}
\end{document}